\documentclass[a4paper,11pt]{article}
\usepackage{jheppub} 
\usepackage{lineno}
\usepackage[normalem]{ulem}
\usepackage{cancel}
\usepackage{xcolor}
\usepackage{float}
\usepackage{appendix}

\newcommand \beq{\begin{eqnarray}}
\newcommand \eeq{\end{eqnarray}} 

\title{From the  universal Lindblad equation to  Boltzmann equations: in-QGP quarkonium dynamics}

\author[a]{Aoumeur Daddi Hammou  }
\author[a]{  Pol Bernard Gossiaux}
\affiliation[a]{SUBATECH, IMT Atlantique, Nantes Université, CNRS/IN2P3,
Nantes, 44307, France}

\emailAdd{daddiham@subatech.in2p3.fr}
\abstract{Recently, a set of coupled singlet--octet universal Lindblad equations (ULEs) was derived within the framework of non-relativistic QCD (NRQCD) to describe quarkonium dynamics in the quark--gluon plasma (QGP). These equations provide a unified quantum description spanning the quantum Brownian and quantum optical regimes. In this work, we further develop this framework and establish its connection with semiclassical transport. We first derive the universal Lindblad equations within the potential non-relativistic QCD (pNRQCD) effective field theory and show that they coincide with the small-dipole limit of the NRQCD ULEs. We then derive the semiclassical limit of the NRQCD ULEs, obtaining a set of coupled singlet--octet Boltzmann equations. To our knowledge, this is the first derivation of Boltzmann transport equations directly from the universal Lindblad framework. The resulting equations are valid beyond the small-dipole approximation, allowing the evolution of heavy-quark pairs from compact to widely separated configurations. Taking their small-dipole limit allows a direct comparison with the Boltzmann equations of Yao \textit{et al.}~[Phys.\ Rev.\ D \textbf{99}, 096028 (2019)], which were derived within pNRQCD from the Davies secular equation, relying on the rotating-wave approximation (RWA). While the singlet equations are found to be in near-complete agreement, the octet equation contains an additional collision term describing transitions within the continuum of octet scattering states, which is absent from the RWA-based derivation. Finally, we derive the leading quantum correction to the singlet Boltzmann equation. Our results establish a more general and systematic theoretical foundation for the semiclassical transport description of quarkonium in the QGP.}

\begin{document}\maketitle
\flushbottom

\section{Introduction}
\label{sec:intro}

Heavy quarkonia are among the primary probes of the quark--gluon plasma (QGP) created in ultra-relativistic heavy-ion collisions (URHICs). Their in-medium evolution is governed by the interplay of screening, dissociation, regeneration, and quantum decoherence, making them particularly sensitive to the microscopic properties of the plasma. Over the past decade, the open quantum systems (OQS) formalism has emerged as a powerful and systematic framework for describing these phenomena from first principles. For recent reviews, see Refs.~\cite{akamatsu2022quarkonium,Yao:2021lus}.

Within this framework, the evolution of a quarkonium state is governed by a quantum master equation. In the Markovian limit, the requirement of complete positivity constrains this equation to take the Lindblad form~\cite{lindblad1976generators,Gorini:1975nb}. Previous derivations of Lindblad equations for in-medium quarkonia assumed a strict hierarchy between the quark--gluon plasma temperature, $T$, and the characteristic quarkonium energy gaps, $\Delta E$, leading to the quantum Brownian ($T\gtrsim\Delta E$) and quantum optical ($\Delta E\gtrsim T$) regimes. Consequently, the resulting Lindblad equations are restricted to their respective domains of applicability. As the QGP expands and cools, the relative magnitude of $T$ and $\Delta E$ evolves, and the quarkonium dynamics may therefore naturally interpolate between these two regimes, motivating a unified description valid throughout the entire evolution.

This limitation was recently overcome in Ref.~\cite{daddigeneral, DaddiHammou:2026,daddi2024}, where the universal Lindblad equation developed by Nathan and Rudner~\cite{nathan2020universal} was applied to the quarkonium--QGP system. The resulting coupled singlet--octet universal Lindblad equations are independent of the quantum Brownian and quantum optical approximations and provide a unified description of in-medium quarkonium dynamics throughout the lifetime of the plasma.

Semiclassical transport equations continue to play a central role in phenomenological studies owing to their relative simplicity and significantly lower computational cost compared with solving the full quantum master equations~\cite{Grandchamp:2003uw,Gossiaux:2004qw,Young:2008he,Zhao:2010nk,Liu:2009nb,Liu:2010ej,Nendzig:2012cu,Zhao:2017yan,Villar:2022sbv,Song:2023zma,Wu:2024gil}. A recent comparison between the quantum and semiclassical descriptions in the quantum Brownian regime, restricted to the Abelian case, showed that the transport equations reproduce the full quantum dynamics with remarkable accuracy~\cite{Daddi-Hammou:2025hdz}. These results provide strong motivation for investigating the semiclassical limit of the universal Lindblad equations and for establishing a controlled connection between the quantum master-equation description and semiclassical transport theory.

In the present work, we derive the semiclassical limit of the coupled singlet--octet universal Lindblad equations obtained in Ref.~\cite{daddigeneral} and obtain the corresponding singlet--octet Boltzmann equations. To our knowledge, this is the first derivation of Boltzmann transport equations directly from the universal Lindblad framework. Since the derivation does not rely on the small-dipole approximation, the resulting transport equations remain valid beyond that limit, providing a more general semiclassical description of quarkonium transport.

This broader range of applicability is particularly relevant for excited quarkonium states, which can have substantially larger spatial extents than the ground states, as well as for stages of the evolution in which the heavy-quark pair reaches large separations. The resulting framework can therefore describe the evolution from compact bound configurations to widely separated $Q\bar Q$ configurations within the same transport description, providing a unified description across the bound and continuum regimes.

We subsequently investigate the small-dipole limit of the derived transport equations and compare them with the Boltzmann equations of Yao \textit{et al.}~\cite{Yao:2018nmy}. The latter were derived within pNRQCD from the Davies secular equation, which relies on the rotating-wave approximation (RWA). The applicability of the RWA to in-medium quarkonium dynamics is, however, not guaranteed. In particular, the continuum of octet scattering states allows transitions with arbitrarily small energy gaps, for which the RWA becomes questionable~\cite{mozgunov2020completely,davidovic2020completely,akamatsu2022quarkonium}. This comparison clarifies the consequences of the rotating-wave approximation and highlights the differences between the resulting transport equations.

The paper is organized as follows. In Sec.~\ref{NRQCD-ULE}, we briefly review the singlet--octet universal Lindblad equations derived in Ref.~\cite{daddigeneral} from the Hamiltonian introduced in Ref.~\cite{blaizot2018quantum}, which, following Refs.~\cite{akamatsu2022quarkonium,Yao:2021lus}, we refer to as the non-relativistic QCD (NRQCD) Hamiltonian. In Sec.~\ref{section-pNRQCD-ULE}, we derive the corresponding universal Lindblad equations within the potential non-relativistic QCD (pNRQCD) framework. In Sec.~\ref{section-NRQCD-Vs-pNRQCD}, we show that they reduce to the small-dipole limit of the NRQCD universal Lindblad equations, establishing the small-dipole limit as the natural framework for comparing the semiclassical transport equations derived from the two effective field theories. In Sec.~\ref{Semiclassical-limit-of-NRQCD-ULEs}, we derive the semiclassical Boltzmann equations, investigate their small-dipole limit, and compare them with the transport equations of Ref.~\cite{Yao:2018nmy}. Finally, in Sec.~\ref{section-quantum-corrections}, we obtain the leading-order quantum correction to the singlet semiclassical equation and compare it with the corresponding result of Ref.~\cite{yao2021semiclassical-transport}. Conclusions and future perspectives are presented in Sec.~\ref{Conclusions}.

\section{NRQCD universal Lindblad equation}
\label{NRQCD-ULE}

In this section, we briefly review the main steps leading to the coupled singlet--octet universal Lindblad equations (ULEs) derived in Ref.~\cite{daddigeneral}. These equations were obtained by applying the universal Lindblad equation developed in Ref.~\cite{nathan2020universal} to the quarkonium--QGP system.

Following the terminology adopted in Refs.~\cite{akamatsu2022quarkonium,Yao:2021lus}, we refer to the Hamiltonian governing the Fock state containing a heavy quark--antiquark pair and employed in Ref.~\cite{daddigeneral} as the NRQCD Hamiltonian. We emphasize that this object is not the full Hamiltonian of NRQCD formulated as an effective field theory. Rather, it is an effective quantum-mechanical Hamiltonian derived from NRQCD by restricting to the heavy-quark sector and retaining the leading interaction between the heavy quark--antiquark pair and the medium. For details of this derivation, we refer the reader to Ref.~\cite{akamatsu2022quarkonium}. The total Hamiltonian is given by \cite{blaizot2018quantum,blaizot2018approach,akamatsu2022quarkonium}
 
\begin{equation}
H_\mathrm{tot}= \left(\frac{\boldsymbol{p}_{Q}^{2}}{2M}+\frac{\boldsymbol{p}_{\bar{Q}}^{2}}{2M}\right)\otimes I_\mathrm{QGP}
+I_{Q\bar{Q}}\otimes H_\mathrm{QGP}
+\int_{\boldsymbol{x}}n_{\boldsymbol{x}}^{a}\otimes gA^a_0(\boldsymbol{x}),
\label{H-NRQCD}
\end{equation}
where the labels $Q$ and $\bar{Q}$ stand for the heavy quark and antiquark, respectively, and $n_{\boldsymbol{x}}^{a}$ is the color charge density defined by
\begin{equation}
    n_{\boldsymbol{x}}^{a}\equiv\delta\left(\boldsymbol{x}-\boldsymbol{x}_Q\right)t^a_Q-\delta\left(\boldsymbol{x}-\boldsymbol{x}_{\bar{Q}}\right)t^{a*}_{\bar{Q}}.
\end{equation} 
The matrices $t^a_Q$ and $t^{a*}_{\bar{Q}}$ are the generators of the color SU($N_c$) algebra in the fundamental representation and its complex conjugate, respectively.  In the last term of Eq.~(\ref{H-NRQCD}), $\boldsymbol{x}$ is just a label, while $\boldsymbol{x}_Q$ and $\boldsymbol{x}_{\bar{Q}}$ are the system operators and the gauge field $A^a_0$ represents the environment operator \cite{blaizot2018quantum, akamatsu2022quarkonium}.

 Using the Hamiltonian (\ref{H-NRQCD}), the NRQCD ULE derived in Ref.~\cite{daddigeneral} can be written as
\begin{equation}
\frac{d\rho\left(t\right)}{dt}=-i\left[H_{Q\bar{Q}}+\Lambda,\rho\left(t\right)\right]+\int_{\boldsymbol{y}}\left(L\left(\boldsymbol{y}\right)\rho\left(t\right)L^{\dagger}\left(\boldsymbol{y}\right)-\frac{1}{2}\left\{ L^{\dagger}\left(\boldsymbol{y}\right)L\left(\boldsymbol{y}\right),\rho\left(t\right)\right\} \right)
\label{eq:ULE}
\end{equation}
where $\rho(t)$ is the density matrix of the reduced system, i.e., the in-QGP quarkonia, and $H_{Q\bar{Q}}$ is the system Hamiltonian.  The Lindblad operator is given by
\begin{equation}\label{eq:Lindblad_operator}
L\left(\boldsymbol{y}\right)=\int_{-\infty}^{+\infty}\text{d}v\int_{\boldsymbol{x}}g\left(t-v,\boldsymbol{y}-\boldsymbol{x}\right)\,U\left(t-v\right)\,n^{a}_{\boldsymbol{x}}\,U^{\dagger}\left(t-v\right),
\end{equation}
and the Lamb-shift term $\Lambda\left(t\right)$ is defined by
\begin{equation}
\begin{split}
\Lambda=&\frac{1}{2i}\int_{-\infty}^{+\infty}\text{d}v\int_{-\infty}^{+\infty}\text{d}v^{\prime}\,\text{sign}\left(v-v^{\prime}\right)\int_{\boldsymbol{y}\boldsymbol{x}\boldsymbol{x}^{\prime}}g\left(v-t,\boldsymbol{x-y}\right)g\left(t-v^{\prime},\boldsymbol{y}-\boldsymbol{x}^{\prime}\right)\\
&\times U\left(t-v\right)\,n^{a}_{\boldsymbol{x}}\,U^{\dagger}\left(t-v\right)\,U\left(t-v^{\prime}\right)\,n^{a}_{\boldsymbol{x}^{\prime}}\,U^{\dagger}\left(t-v^{\prime}\right),
\end{split}\label{eq:24}
\end{equation}
with 
$U\left(t-v\right)=e^{-iH_{Q}\left(t-v\right)}$ being the evolution operator.

The so-called ``jump correlator function'' $g(t,\boldsymbol{x})$ is related to the QGP thermal correlator $\Delta^{>}\left(t,\boldsymbol{x}\right)$   through \cite{nathan2020universal,daddigeneral}
\begin{equation}
\Delta^{>}\left(t-t^{\prime},\boldsymbol{x}-\boldsymbol{x}^{\prime}\right)=\int_{-\infty}^{+\infty}\text{d}v\int_{\boldsymbol{y}}g\left(t-v,\boldsymbol{x-y}\right)\,g\left(v-t^{\prime},\boldsymbol{y}-\boldsymbol{x}^{\prime}\right),\label{convolution-ULE}
\end{equation}
with 
\begin{align}\label{thermal-propagators}
\delta^{ab}\Delta^{>}\left(t_1-t_2,\boldsymbol{x}-\boldsymbol{x}^{\prime}\right)&=
g^{2}\left\langle T_{C}\left[A_{0}^{a}\left(t_{1},\boldsymbol{x}\right)A_{0}^{b}\left(t_{2},\boldsymbol{x}^{\prime}\right)\right]\right\rangle _{0},\\
\Delta^{<}\left(t_1-t_2,\boldsymbol{x}-\boldsymbol{x}^{\prime}\right)&=\Delta^{>}\left(t_2-t_1,\boldsymbol{x}-\boldsymbol{x}^{\prime}\right),\notag
\end{align}
where $T_c$ is the time order operator along the Schwinger-Keldysh contour, and  $\langle \cdots\rangle_0$ denotes the  average with respect to the QGP equilibrium density matrix \cite{blaizot2018quantum}.

Using the Fourier transform of the QGP correlator,  the jump correlator can be defined as follows:
\begin{equation}
g\left(t\right)=\frac{1}{2\pi}\int\text{d}q_{0}\, e^{-iq_{0} t}\sqrt{\Delta^{>}\left(q_{0}\right)}, \label{g-as-function-of-Delta}
\end{equation}
and its complex conjugate satisfies $g^*\left(t\right)=g\left(-t\right)$.\footnote{A similar definition can be obtained with respect to the position sector, see \cite{daddigeneral}.} It can be shown that in Fourier space, the following relation holds
\begin{equation}
\Delta^{>}\left(q_{0}\right)=\left(g\left(q_{0}\right)\right)^{2}\label{eq:10}.
\end{equation}

It is worth noting that the jump correlator and the QGP thermal correlator are characterized by the same decay time scale, namely, the medium correlation time $\tau_E$. 

After projecting onto color space and tracing out the center-of-mass degrees of freedom, the coupled singlet--octet ULEs can be expressed in an explicit Lindblad form; see Ref.~\cite{daddigeneral}. For the comparison with the pNRQCD ULEs and the subsequent derivation of the Boltzmann equations, it is convenient to rewrite them in a form similar to that adopted in Ref.~\cite{blaizot2018approach}
\footnote{The tilde in these expressions  refers to the operators after integrating out the center of mass, see \cite{blaizot2018approach,daddi2024}.}  
 \begin{equation}
\frac{d\tilde{\rho}_{s}\left(t\right)}{dt}=-i\left[\tilde{H}_{s}+\tilde{\Lambda}_{s},\tilde{\rho}_{s}\left(t\right)\right]+\tilde{\mathcal{L}}_{ss}\left(t\right)\tilde{\rho_{s}}\left(t\right)+\mathcal{\tilde{L}}_{so}\left(t\right)\tilde{\rho}_{o}\left(t\right)\label{eq:sinlget-equation-com-integrated-out},
\end{equation}
\begin{equation}
\frac{d\tilde{\rho}_{o}\left(t\right)}{dt}=-i\left[\tilde{H}_{o}+\tilde{\Lambda}_{o},\tilde{\rho}_{o}\left(t\right)\right]+\tilde{\mathcal{L}}_{os}\left(t\right)\tilde{\rho_{s}}\left(t\right)+\mathcal{\tilde{L}}_{oo}\left(t\right)\tilde{\rho}_{o}\left(t\right)\label{eq:octet-equation-com-integrated-out},
\end{equation}
where the first equation corresponds to the singlet ULE and the second one to the octet ULE. The Liouville superoperators  $\tilde{\mathcal{L}}_{so}$ and $\tilde{\mathcal{L}}_{os}$ describe dissociation and regeneration, respectively. 

The expressions of the  Lamb-shift and Liouville superoperators  of the singlet ULE are given by \footnote{ The computational details underlying the derivation of these expressions can be found in \cite{daddi2024}.}
\begin{equation}
\begin{array}{ccl}
\tilde{\Lambda}_{s}& = & \frac{C_{F}}{2i}\int_{vv^{\prime}q_{0}q_{0}^{\prime}\boldsymbol{q}}\text{sign}\left(v-v^{\prime}\right)e^{i\left(q_{0}-q_{0}^{\prime}\right)t}e^{-iq_{0}v+iq_{0}^{\prime}v^{\prime}}g\left(q_{0},\boldsymbol{q}\right)g\left(q_{0}^{\prime},\boldsymbol{q}\right)\\
 &  & \times\tilde{U}_{s}\left(t-v\right)S_{\boldsymbol{q}.\hat{\boldsymbol{s}}}\tilde{U}_{o}^{\dagger}\left(t-v\right)\tilde{U}_{o}\left(t-v^{\prime}\right)S_{\boldsymbol{q}.\hat{\boldsymbol{s}}^{\prime}}\tilde{U}_{s}^{\dagger}\left(t-v^{\prime}\right)
\end{array}
\end{equation}
\begin{equation}
\begin{array}{ccl}
\tilde{\mathcal{L}}_{ss}\left(t\right)\tilde{\rho_{s}}\left(t\right) & = & \frac{-C_{F}}{2}\int_{vv^{\prime}q_{0}q_{0}^{\prime}\boldsymbol{q}}e^{i\left(q_{0}-q_{0}^{\prime}\right)t}e^{-iq_{0}v+iq_{0}^{\prime}v^{\prime}}g\left(q_{0},\boldsymbol{q}\right)g\left(q_{0}^{\prime},\boldsymbol{q}\right)\\
 &  & \times\left\{ \tilde{U}_{s}\left(t-v\right)S_{\boldsymbol{q}.\hat{\boldsymbol{s}}}\tilde{U}_{o}^{\dagger}\left(t-v\right)\tilde{U}_{o}\left(t-v^{\prime}\right)S_{\boldsymbol{q}.\hat{\boldsymbol{s}}^{\prime}}\tilde{U}_{s}^{\dagger}\left(t-v^{\prime}\right),\tilde{\rho}_{s}\left(t\right)\right\} 
\end{array}\label{eq:121}
\end{equation}
\begin{equation}
\begin{array}{ccl}
\tilde{\mathcal{L}}_{so}\left(t\right)\tilde{\rho_{o}}\left(t\right) & = & C_{F}\int_{vv^{\prime}q_{0}q_{0}^{\prime}\boldsymbol{q}}e^{i\left(q_{0}-q_{0}^{\prime}\right)t}e^{-iq_{0}v+iq_{0}^{\prime}v^{\prime}}g\left(q_{0},\boldsymbol{q}\right)g\left(q_{0}^{\prime},\boldsymbol{q}\right)\\
 &  & \times\tilde{U}_{s}\left(t-v^{\prime}\right)S_{\boldsymbol{q}.\hat{\boldsymbol{s}}^{\prime}}\tilde{U}_{o}^{\dagger}\left(t-v^{\prime}\right)\tilde{\rho}_{o}\left(t\right)\tilde{U}_{o}\left(t-v\right)S_{\boldsymbol{q}.\hat{\boldsymbol{s}}}\tilde{U}_{s}^{\dagger}\left(t-v\right)
\end{array}\label{eq:148}
\end{equation}

Similarly, expressions of the  Lamb-shift and Liouville superoperators  of the octet  ULE are given by:
\begin{equation}
\begin{array}{ccl}
\tilde{\Lambda}_{o}  & = & \frac{1}{8i}\frac{N_{c}^{2}-4}{N_{c}}\int_{vv^{\prime}q_{0}q_{0}^{\prime}\boldsymbol{q}}\text{sign}\left(v-v^{\prime}\right)e^{i\left(q_{0}-q_{0}^{\prime}\right)t}e^{-iq_{0}v+iq_{0}^{\prime}v^{\prime}}g\left(q_{0},\boldsymbol{q}\right)g\left(q_{0}^{\prime},\boldsymbol{q}\right)\\
 &  & \times\tilde{U}_{o}\left(t-v\right)S_{\boldsymbol{q}.\hat{\boldsymbol{s}}}\tilde{U}_{o}^{\dagger}\left(t-v\right)\tilde{U}_{o}\left(t-v^{\prime}\right)S_{\boldsymbol{q}.\hat{\boldsymbol{s}}^{\prime}}\tilde{U}_{o}^{\dagger}\left(t-v^{\prime}\right)\\
\\
 &  & +\frac{N_{c}}{8i}\int_{vv^{\prime}q_{0}q_{0}^{\prime}\boldsymbol{q}}\text{sign}\left(v-v^{\prime}\right)e^{i\left(q_{0}-q_{0}^{\prime}\right)t}e^{-iq_{0}v+iq_{0}^{\prime}v^{\prime}}g\left(q_{0},\boldsymbol{q}\right)g\left(q_{0}^{\prime},\boldsymbol{q}\right)\\
 &  & \times\tilde{U}_{o}\left(t-v\right)C_{\boldsymbol{q}.\hat{\boldsymbol{s}}}\tilde{U}_{o}^{\dagger}\left(t-v\right)\tilde{U}_{o}\left(t-v^{\prime}\right)C_{\boldsymbol{q}.\hat{\boldsymbol{s}}^{\prime}}\tilde{U}_{o}^{\dagger}\left(t-v^{\prime}\right)\\
\\
 &  & +\frac{1}{4iN_{c}}\int_{vv^{\prime}q_{0}q_{0}^{\prime}\boldsymbol{q}}\text{sign}\left(v-v^{\prime}\right)e^{i\left(q_{0}-q_{0}^{\prime}\right)t}e^{-iq_{0}v+iq_{0}^{\prime}v^{\prime}}g\left(q_{0},\boldsymbol{q}\right)g\left(q_{0}^{\prime},\boldsymbol{q}\right)\\
 &  & \times\tilde{U}_{o}\left(t-v\right)S_{\boldsymbol{q}.\hat{\boldsymbol{s}}}\tilde{U}_{s}^{\dagger}\left(t-v\right)\tilde{U}_{s}\left(t-v^{\prime}\right)S_{\boldsymbol{q}.\hat{\boldsymbol{s}}^{\prime}}\tilde{U}_{o}^{\dagger}\left(t-v^{\prime}\right)\\
\end{array}\label{eq:125},
\end{equation}
\begin{equation}
\begin{array}{ccl}
\tilde{\mathcal{L}}_{os}\left(t\right)\tilde{\rho}_{s}\left(t\right) & = & \frac{1}{2N_{c}}\int_{vv^{\prime}q_{0}q_{0}^{\prime}\boldsymbol{q}}e^{i\left(q_{0}-q_{0}^{\prime}\right)t}e^{-iq_{0}v+iq_{0}^{\prime}v^{\prime}}g\left(q_{0},\boldsymbol{q}\right)g\left(q_{0}^{\prime},\boldsymbol{q}\right)\\
 &  & \times\tilde{U}_{o}\left(t-v^{\prime}\right)S_{\boldsymbol{q}.\hat{\boldsymbol{s}}^{\prime}}\tilde{U}_{s}^{\dagger}\left(t-v^{\prime}\right)\tilde{\rho}_{s}\left(t\right)\tilde{U}_{s}\left(t-v\right)S_{\boldsymbol{q}.\hat{\boldsymbol{s}}}\tilde{U}_{o}^{\dagger}\left(t-v\right)\\
\end{array}\label{eq:126},
\end{equation}

\begin{equation}
\begin{array}{ccl}
\tilde{\mathcal{L}}_{oo}\left(t\right)\tilde{\rho}_{o}\left(t\right) & = & \frac{N_{c}^{2}-4}{4N_{c}}\int_{vv^{\prime}q_{0}q_{0}^{\prime}\boldsymbol{q}}e^{i\left(q_{0}-q_{0}^{\prime}\right)t}e^{-iq_{0}v+iq_{0}^{\prime}v^{\prime}}g\left(q_{0},\boldsymbol{q}\right)g\left(q_{0}^{\prime},\boldsymbol{q}\right)\\
 &  & \times\tilde{U}_{o}\left(t-v^{\prime}\right)S_{\boldsymbol{q}.\hat{\boldsymbol{s}}^{\prime}}\tilde{U}_{o}^{\dagger}\left(t-v^{\prime}\right)\tilde{\rho}_{o}\left(t\right)\tilde{U}_{o}\left(t-v\right)S_{\boldsymbol{q}.\hat{\boldsymbol{s}}}\tilde{U}_{o}^{\dagger}\left(t-v\right)\\
 \\
 &  & +\frac{N_{c}}{4}\int_{vv^{\prime}q_{0}q_{0}^{\prime}\boldsymbol{q}}e^{i\left(q_{0}-q_{0}^{\prime}\right)t}e^{-iq_{0}v+iq_{0}^{\prime}v^{\prime}}g\left(q_{0},\boldsymbol{q}\right)g\left(q_{0}^{\prime},\boldsymbol{q}\right)\\
 &  & \times\tilde{U}_{o}\left(t-v^{\prime}\right)C_{\boldsymbol{q}.\hat{\boldsymbol{s}}^{\prime}}\tilde{U}_{o}^{\dagger}\left(t-v^{\prime}\right)\tilde{\rho}_{o}\left(t\right)\tilde{U}_{o}\left(t-v\right)C_{\boldsymbol{q}.\hat{\boldsymbol{s}}}\tilde{U}_{o}^{\dagger}\left(t-v\right)\\
\\
 &  & -\frac{1}{4N_{c}}\int_{vv^{\prime}q_{0}q_{0}^{\prime}\boldsymbol{q}}e^{i\left(q_{0}-q_{0}^{\prime}\right)t}e^{-iq_{0}v+iq_{0}^{\prime}v^{\prime}}g\left(q_{0},\boldsymbol{q}\right)g\left(q_{0}^{\prime},\boldsymbol{q}\right)\\
 &  & \times\left\{ \tilde{U}_{o}\left(t-v\right)S_{\boldsymbol{q}.\hat{\boldsymbol{s}}}\tilde{U}_{s}^{\dagger}\left(t-v\right)\tilde{U}_{s}\left(t-v^{\prime}\right)S_{\boldsymbol{q}.\hat{\boldsymbol{s}}^{\prime}}\tilde{U}_{o}^{\dagger}\left(t-v^{\prime}\right),\tilde{\rho}_{o}\left(t\right)\right\} \\
\\ &  & -\frac{N_{c}^2-4}{8N_c}\int_{vv^{\prime}q_{0}q_{0}^{\prime}\boldsymbol{q}}e^{i\left(q_{0}-q_{0}^{\prime}\right)t}e^{-iq_{0}v+iq_{0}^{\prime}v^{\prime}}g\left(q_{0},\boldsymbol{q}\right)g\left(q_{0}^{\prime},\boldsymbol{q}\right)\\
 &  & \times\left\{ \tilde{U}_{o}\left(t-v\right)S_{\boldsymbol{q}.\hat{\boldsymbol{s}}}\tilde{U}_{o}^{\dagger}\left(t-v\right)\tilde{U}_{o}\left(t-v^{\prime}\right)S_{\boldsymbol{q}.\hat{\boldsymbol{s}}^{\prime}}\tilde{U}_{o}^{\dagger}\left(t-v^{\prime}\right),\tilde{\rho}_{o}\left(t\right)\right\}\\
\\
 &  & -\frac{N_{c}}{8}\int_{vv^{\prime}q_{0}q_{0}^{\prime}\boldsymbol{q}}e^{i\left(q_{0}-q_{0}^{\prime}\right)t}e^{-iq_{0}v+iq_{0}^{\prime}v^{\prime}}g\left(q_{0},\boldsymbol{q}\right)g\left(q_{0}^{\prime},\boldsymbol{q}\right)\\
 &  & \times\left\{ \tilde{U}_{o}\left(t-v\right)C_{\boldsymbol{q}.\hat{\boldsymbol{s}}}\tilde{U}_{o}^{\dagger}\left(t-v\right)\tilde{U}_{o}\left(t-v^{\prime}\right)C_{\boldsymbol{q}.\hat{\boldsymbol{s}}^{\prime}}\tilde{U}_{o}^{\dagger}\left(t-v^{\prime}\right),\tilde{\rho}_{o}\left(t\right)\right\} \\
\end{array}\label{eq:127},
\end{equation}

In these equations, the functions $S_{\boldsymbol{q}.\hat{\boldsymbol{s}}}$ and $C_{\boldsymbol{q}.\hat{\boldsymbol{s}}}$ are defined as follows:
\begin{equation}
 S_{\boldsymbol{q}.\hat{\boldsymbol{s}}}\equiv2\text{sin}\left(\frac{\boldsymbol{q}.\hat{\boldsymbol{s}}}{2}\right), \label{eq:2.17}  
\end{equation}
\begin{equation}
C_{\boldsymbol{q}.\hat{\boldsymbol{s}}}\equiv2\text{cos}\left(\frac{\boldsymbol{q}.\hat{\boldsymbol{s}}}{2}\right),\label{eq:2.18} 
\end{equation}
where $\hat{\boldsymbol{s}}$ is the relative distance operator between the heavy quark pair. 

The expressions above will serve as the starting point for the comparison with the corresponding pNRQCD equations in the next section. In particular, this comparison will establish the relation between the two formulations in the small-dipole limit, which will subsequently provide the appropriate framework for comparing the semiclassical transport equations.

\section{pNRQCD universal Lindblad equation}
\label{section-pNRQCD-ULE}

We now apply the universal Lindblad equation to the quarkonium--QGP system within the pNRQCD framework and derive the corresponding coupled singlet--octet equations. To this end, we use the pNRQCD Hamiltonian~\cite{akamatsu2022quarkonium} 
\footnote{We adhere to the conventional notation and denote the dipole size by $\boldsymbol{r}$. In the NRQCD ULEs, the same quantity is denoted by $\boldsymbol{s}$; see the remark below Eq.~(\ref{eq:2.18}).}
\begin{equation}
\begin{split}
 H_\mathrm{tot}=&\frac{\boldsymbol{p}^{2}}{M}-\frac{C_{F}\alpha_{s}}{\boldsymbol{r}}\left|s\right\rangle \left\langle s\right|+\frac{\alpha_{s}}{2N_{c}\boldsymbol{r}}\left|a\right\rangle \left\langle a\right|\otimes I_\mathrm{QGP}+I_{Q\bar{Q}}\otimes H_\mathrm{QGP}\\
 &-r_{i}\left[\sqrt{\frac{1}{2N_{c}}}\left(\left|a\right\rangle \left\langle s\right|+\left|s\right\rangle \left\langle a\right|\right)+\frac{1}{2}d^{abc}\left|b\right\rangle \left\langle c\right|\right]\otimes g\tilde{E}_{i}^{a}\left(\boldsymbol{R}\right),
\end{split}
\label{eq:176}
\end{equation}
where $r_i$ is the dipole-size operator, $\tilde{E}_{i}^{a}$ is the chromoelectric field, and $\boldsymbol{R}$ is the center-of-mass coordinate; see, e.g., Refs.~\cite{akamatsu2022quarkonium,Yao:2021lus} and references therein for further details.

Following the same procedure used to derive the NRQCD ULE in Ref.~\cite{daddigeneral}, we obtain the pNRQCD ULE
\begin{equation}
\frac{d\rho\left(t\right)}{dt}=-i\left[H_{Q\bar{Q}}+\Lambda,\rho\left(t\right)\right]+\int_{\boldsymbol{y}}\left(L\left(\boldsymbol{y}\right)\rho\left(t\right)L^{\dagger}\left(\boldsymbol{y}\right)-\frac{1}{2}\left\{ L^{\dagger}\left(\boldsymbol{y}\right)L\left(\boldsymbol{y}\right),\rho\left(t\right)\right\} \right)
\label{eq:22-1}
\end{equation}
where the Lindblad operator is given by
\begin{equation}
L_{i}\left(\boldsymbol{y}\right)=\int_{-\infty}^{+\infty}\text{d}vg\left(t-v,\boldsymbol{y}-\boldsymbol{R}\right)U\left(t-v\right)V_{i}^{a}U^{\dagger}\left(t-v\right),
\label{eq:23-1}
\end{equation}
and the Lamb-shift term is
\begin{equation}
\Lambda=\frac{1}{2i}\int_{vv^{\prime}\boldsymbol{y}}\text{sign}\left(v-v^{\prime}\right)g\left(v-t,\boldsymbol{R}-\boldsymbol{y}\right)g\left(t-v^{\prime},\boldsymbol{y}-\boldsymbol{R}\right)U\left(t-v\right)V_{i}^{a}U^{\dagger}\left(t-v\right)U\left(t-v^{\prime}\right)V_{i}^{a}U^{\dagger}\left(t-v^{\prime}\right)
\label{eq:24-1}
\end{equation}
with
\begin{equation}
V_{i}^{a}\equiv r_{i}\left[\sqrt{\frac{1}{2N_{c}}}\left(\left|a\right\rangle \left\langle s\right|+\left|s\right\rangle \left\langle a\right|\right)+\frac{1}{2}d^{abc}\left|b\right\rangle \left\langle c\right|\right]
\end{equation}
and
\begin{equation}
\int_{vv^{\prime}\boldsymbol{y}}\equiv\int_{-\infty}^{+\infty}\text{d}v\int_{-\infty}^{+\infty}\text{d}v^{\prime}\int_{-\infty}^{+\infty}\text{d}\boldsymbol{y}.
\end{equation}

The jump correlator $g(t,\boldsymbol{x})$ that appears in the pNRQCD ULE is defined by
\footnote{The delta function $\delta_{ij}$ originates from the isotropy of the QGP.}
\begin{equation}
g^{2}Tr_\mathrm{QGP}\left(\rho_\mathrm{QGP}\tilde{E}_{i}^{a}\left(t,\boldsymbol{R}\right)\tilde{E}_{j}^{b}\left(t^{\prime},\boldsymbol{R}^{\prime}\right)\right)=\delta_{ij}\delta^{ab}\int_{-\infty}^{+\infty}\int_{-\infty}^{+\infty}\text{d}v\text{d}\boldsymbol{y}\,g\left(t-v,\boldsymbol{R}-\boldsymbol{y}\right)g\left(v-t^{\prime},\boldsymbol{y}-\boldsymbol{R}^{\prime}\right).
\label{eq:177}
\end{equation}

The jump correlator $g(t,\boldsymbol{x})$ appearing in the NRQCD ULE (\ref{eq:ULE}) differs from that appearing in the pNRQCD ULE (\ref{eq:22-1}).  In the former, $g(t,\boldsymbol{x})$ is defined in terms of the correlator of Coulombic gluons, $A_{0}^{a}(t,\boldsymbol{x})$, whereas in the latter it is defined in terms of the correlator of chromoelectric fields, $\tilde{E}_{i}^{a}(t,\boldsymbol{R})$. This distinction will be important when comparing the singlet--octet ULEs derived within the NRQCD and pNRQCD frameworks.

For the comparison below, it is useful to express the pNRQCD coupled singlet--octet ULEs in a form analogous to the NRQCD singlet--octet ULEs, Eqs.~(\ref{eq:sinlget-equation-com-integrated-out}) and (\ref{eq:octet-equation-com-integrated-out}). After projection onto color space, the pNRQCD singlet--octet ULEs can be written as
\footnote{We neglect the recoil energy $\Delta E$ in this work, i.e., $\Delta E\simeq0$. Including this energy in the singlet--octet ULEs is straightforward. In that case, however, solving the ULEs also requires treating the center-of-mass dynamics.}
\begin{equation}
\frac{d\tilde{\rho}_{s}\left(t\right)}{dt}=-i\left[H_{s}+\tilde{\Lambda}_{s},\tilde{\rho}_{s}\left(t\right)\right]+\tilde{\mathcal{L}}_{ss}\left(t\right)\tilde{\rho}_{s}\left(t\right)+\tilde{\mathcal{L}}_{so}\left(t\right)\tilde{\rho}_{o}\left(t\right)\label{eq:singlet equation22},
\end{equation}
\begin{equation}
\frac{d\tilde{\rho}_{o}\left(t\right)}{dt}=-i\left[H_{o}+\tilde{\Lambda}_{o},\tilde{\rho}_{o}\left(t\right)\right]+\tilde{\mathcal{L}}_{os}\left(t\right)\tilde{\rho}_{s}\left(t\right)+\tilde{\mathcal{L}}_{oo}\left(t\right)\tilde{\rho}_{o}\left(t\right)\label{octet eqaution22}.
\end{equation}

The singlet Lamb-shift and Liouville superoperators are given by
\begin{equation}
\begin{array}{ccl}
\tilde{\Lambda}_{s}  & = & \frac{C_{F}}{2i}\int_{vv^{\prime}q_{0}q_{0}^{\prime}\boldsymbol{q}}\text{sign}\left(v-v^{\prime}\right)e^{i\left(q_{0}-q_{0}^{\prime}\right)t}e^{-iq_{0}v+iq_{0}^{\prime}v^{\prime}}g\left(q_{0},\boldsymbol{q}\right)g\left(q_{0}^{\prime},\boldsymbol{q}\right)\\ 
 && \times \tilde{U}_{s}\left(t-v\right)r_{i}\tilde{U}_{o}^{\dagger}\left(t-v\right)\tilde{U}_{o}\left(t-v^{\prime}\right)r_{i}^{\prime}\tilde{U}_{s}^{\dagger}\left(t-v^{\prime}\right)
\end{array},
\end{equation}
\begin{equation}
\begin{array}{ccl}
\tilde{\mathcal{L}}_{ss}\left(t\right)\tilde{\rho}_{s}\left(t\right) & = & -\frac{C_{F}}{2}\int_{vv^{\prime}q_{0}q_{0}^{\prime}\boldsymbol{q}}e^{i\left(q_{0}-q_{0}^{\prime}\right)t}e^{-iq_{0}v+iq_{0}^{\prime}v^{\prime}}g\left(q_{0},\boldsymbol{q}\right)g\left(q_{0}^{\prime},\boldsymbol{q}\right)\\
&& \times \left\{ \tilde{U}_{s}\left(t-v\right)r_{i}\tilde{U}_{o}^{\dagger}\left(t-v\right)\tilde{U}_{o}\left(t-v^{\prime}\right)r_{i}^{\prime}\tilde{U}_{s}^{\dagger}\left(t-v^{\prime}\right),\tilde{\rho}_{s}\left(t\right)\right\} 
\end{array}\label{eq:187}\\,
\end{equation}
\begin{equation}
\begin{array}{ccl}
\tilde{\mathcal{L}}_{so}\left(t\right)\tilde{\rho}_{o}\left(t\right) & = & C_{F}\int_{vv^{\prime}q_{0}q_{0}^{\prime}\boldsymbol{q}}e^{i\left(q_{0}-q_{0}^{\prime}\right)t}e^{-iq_{0}v+iq_{0}^{\prime}v^{\prime}}g\left(q_{0},\boldsymbol{q}\right)g\left(q_{0}^{\prime},\boldsymbol{q}\right)\\
&& \times \tilde{U}_{s}\left(t-v\right)r_{i}\tilde{U}_{o}^{\dagger}\left(t-v\right)\tilde{\rho}_{o}\left(t\right)\tilde{U}_{o}\left(t-v^{\prime}\right)r_{i}^{\prime}\tilde{U}_{s}^{\dagger}\left(t-v^{\prime}\right)
\end{array}.
\end{equation}

The details of the color projection leading to these expressions are given in Appendix~\ref{Appendinx-pNRQCD-Color-Projection}.

Similarly, the octet Lamb-shift and Liouville superoperators are given by
\begin{equation}
\protect\begin{array}{ccl}
\protect\\
\tilde{\Lambda}_{o}  & = & \frac{1}{4iN_{c}}\int_{vv^{\prime}q_{0}q_{0}^{\prime}\boldsymbol{q}}\text{sign}\left(v-v^{\prime}\right)e^{i\left(q_{0}-q_{0}^{\prime}\right)t}e^{-iq_{0}v+iq_{0}^{\prime}v^{\prime}}g\left(q_{0},\boldsymbol{q}\right)g\left(q_{0}^{\prime},\boldsymbol{q}\right)\\
&& \times \tilde{U}_{o}\left(t-v\right)r_{i}\tilde{U}_{s}^{\dagger}\left(t-v\right)\tilde{U}_{s}\left(t-v^{\prime}\right)r_{i}^{\prime}\tilde{U}_{o}^{\dagger}\left(t-v^{\prime}\right)\\
\\
 &  & +\frac{N_{c}^{2}-4}{8iN_{c}}\int_{vv^{\prime}q_{0}q_{0}^{\prime}\boldsymbol{q}}\text{sign}\left(v-v^{\prime}\right)e^{i\left(q_{0}-q_{0}^{\prime}\right)t}e^{-iq_{0}v+iq_{0}^{\prime}v^{\prime}}g\left(q_{0},\boldsymbol{q}\right)g\left(q_{0}^{\prime},\boldsymbol{q}\right)\\
 && \times \tilde{U}_{o}\left(t-v\right)r_{i}\tilde{U}_{o}^{\dagger}\left(t-v\right)\tilde{U}_{o}\left(t-v^{\prime}\right)r_{i}^{\prime}\tilde{U}_{o}^{\dagger}\left(t-v^{\prime}\right)
\end{array}\label{pNRQCD-octet-LS},
\end{equation}
\begin{equation}
\protect\begin{array}{ccl}
\protect
\tilde{\mathcal{L}}_{os}\left(t\right)\tilde{\rho}_{s}\left(t\right) & = & \frac{1}{2N_{c}}\int_{vv^{\prime}q_{0}q_{0}^{\prime}\boldsymbol{q}}e^{i\left(q_{0}-q_{0}^{\prime}\right)t}e^{-iq_{0}v+iq_{0}^{\prime}v^{\prime}}g\left(q_{0},\boldsymbol{q}\right)g\left(q_{0}^{\prime},\boldsymbol{q}\right)\\
&& \times \tilde{U}_{o}\left(t-v\right)r_{i}\tilde{U}_{s}^{\dagger}\left(t-v\right)\tilde{\rho}_{s}\left(t\right)\tilde{U}_{s}\left(t-v^{\prime}\right)r_{i}^{\prime}\tilde{U}_{o}^{\dagger}\left(t-v^{\prime}\right)\protect
\end{array}\label{eq:206},\\
\end{equation}
\begin{equation}
\\
\protect\begin{array}{ccl}
\protect
\tilde{\mathcal{L}}_{oo}\left(t\right)\tilde{\rho}_{o}\left(t\right) & = & \frac{N_{c}^{2}-4}{4N_{c}}\int_{vv^{\prime}q_{0}q_{0}^{\prime}\boldsymbol{q}}e^{i\left(q_{0}-q_{0}^{\prime}\right)t}e^{-iq_{0}v+iq_{0}^{\prime}v^{\prime}}g\left(q_{0},\boldsymbol{q}\right)g\left(q_{0}^{\prime},\boldsymbol{q}\right)\\
&& \times \tilde{U}_{o}\left(t-v\right)r_{i}\tilde{U}_{o}^{\dagger}\left(t-v\right)\tilde{\rho}_{o}\left(t\right)\tilde{U}_{o}\left(t-v^{\prime}\right)r_{i}^{\prime}\tilde{U}_{o}^{\dagger}\left(t-v^{\prime}\right)\protect\\
\\
 &  & -\frac{1}{4N_{c}}\int_{vv^{\prime}q_{0}q_{0}^{\prime}\boldsymbol{q}}e^{i\left(q_{0}-q_{0}^{\prime}\right)t}e^{-iq_{0}v+iq_{0}^{\prime}v^{\prime}}g\left(q_{0},\boldsymbol{q}\right)g\left(q_{0}^{\prime},\boldsymbol{q}\right)\\
 && \times \left\{ \tilde{U}_{o}\left(t-v\right)r_{i}\tilde{U}_{s}^{\dagger}\left(t-v\right)\tilde{U}_{s}\left(t-v^{\prime}\right)r_{i}^{\prime}\tilde{U}_{o}^{\dagger}\left(t-v^{\prime}\right),\tilde{\rho}_{o}\left(t\right)\right\} \\
 \\
 &  & -\frac{N_{c}^{2}-4}{8N_{c}}\int_{vv^{\prime}q_{0}q_{0}^{\prime}\boldsymbol{q}}e^{i\left(q_{0}-q_{0}^{\prime}\right)t}e^{-iq_{0}v+iq_{0}^{\prime}v^{\prime}}g\left(q_{0},\boldsymbol{q}\right)g\left(q_{0}^{\prime},\boldsymbol{q}\right)\\
&& \times\left\{ \tilde{U}_{o}\left(t-v\right)r_{i}\tilde{U}_{o}^{\dagger}\left(t-v\right)\tilde{U}_{o}\left(t-v^{\prime}\right)r_{i}^{\prime}\tilde{U}_{o}^{\dagger}\left(t-v^{\prime}\right),\tilde{\rho}_{o}\left(t\right)\right\} 
\end{array}\label{pNRQCD-octet-octet-element}.
\end{equation}

\section{NRQCD vs. pNRQCD ULEs}\label{section-NRQCD-Vs-pNRQCD}

In this section, we compare the pNRQCD singlet--octet ULEs with the small-dipole limit of the corresponding NRQCD ULEs. This comparison establishes the correspondence between the two effective-field-theory descriptions and, in particular, shows that the small-dipole limit provides the appropriate framework for comparing the semiclassical transport equations in the following section. A related comparison was performed in Ref.~\cite{akamatsu2022quarkonium} for quantum Brownian master equations. Here, we extend this analysis to the universal Lindblad framework, which applies to both the quantum Brownian and quantum optical regimes.
The small-dipole expansion of Eqs.~(\ref{eq:sinlget-equation-com-integrated-out}) and (\ref{eq:octet-equation-com-integrated-out}) can be performed directly at the level of the functions $S_{\boldsymbol{q}.\hat{\boldsymbol{s}}}$ and $C_{\boldsymbol{q}.\hat{\boldsymbol{s}}}$. Expanding these functions in powers of the dipole size, $\boldsymbol{s}\ll1$, gives at leading order
\begin{equation}
S_{\boldsymbol{q}.\hat{\boldsymbol{s}}}\approx\boldsymbol{q}.\hat{\boldsymbol{s}},\,\text{and}\,\,\,C_{\boldsymbol{q}.\hat{\boldsymbol{s}}}\approx2\hat{I}\label{eq:small-s-limit},
\end{equation}
where $\hat{I}$ is the identity operator.\footnote{Notably, during the course of this project, the expansion given by Eqs.~(\ref{eq:small-s-limit}) was independently introduced and implemented in Ref.~\cite{Mtz-Vera:2024aom}, with the objective of extending the applicability of the master equation developed in Refs.~\cite{brambilla2021bottomonium,Brambilla:2022ynh}.}

\subsection{NRQCD vs. pNRQCD singlet ULE}

The small-dipole limit of the NRQCD singlet equation (\ref{eq:sinlget-equation-com-integrated-out}) is obtained using Eq.~(\ref{eq:small-s-limit}). The resulting expression can then be compared with the singlet pNRQCD master equation derived in the previous section.

We consider the contribution $\mathcal{L}_{ss}\left(t\right)\rho_s\left(t\right)$; the same argument applies to the other contributions entering the singlet ULE in Eq.~(\ref{eq:sinlget-equation-com-integrated-out}).

The small-dipole limit of the NRQCD expression for this contribution is
\begin{equation}
\begin{array}{ccl}
\mathcal{L}_{ss}^\mathrm{NRQCD}\left(t\right)\rho_{s}\left(t\right) & = & \frac{-C_{F}}{2}\int_{vv^{\prime}q_{0}q_{0}^{\prime}\boldsymbol{q}}e^{i\left(q_{0}-q_{0}^{\prime}\right)t}e^{-iq_{0}v+iq_{0}^{\prime}v^{\prime}}g\left(q_{0},\boldsymbol{q}\right)_\mathrm{NRQCD}g\left(q_{0}^{\prime},\boldsymbol{q}\right)_\mathrm{NRQCD}\\
 &  & \times\left\{ \tilde{U}_{s}\left(t-v\right)\boldsymbol{q}.\hat{\boldsymbol{s}}\tilde{U}_{o}^{\dagger}\left(t-v\right)\tilde{U}_{o}\left(t-v^{\prime}\right)\boldsymbol{q}.\hat{\boldsymbol{s}}^{\prime}\tilde{U}_{s}^{\dagger}\left(t-v^{\prime}\right),\tilde{\rho}_{s}\left(t\right)\right\} \\
\end{array}\label{eq:191},
\end{equation}
which can be compared with the corresponding pNRQCD expression,
\begin{equation}
\begin{array}{ccl}
\\
\mathcal{L}_{ss}^\mathrm{pNRQCD}\left(t\right)\rho_{s}\left(t\right) & = & \frac{-C_{F}}{2}\int_{vv^{\prime}q_{0}q_{0}^{\prime}\boldsymbol{q}}e^{i\left(q_{0}-q_{0}^{\prime}\right)t}e^{-iq_{0}v+iq_{0}^{\prime}v^{\prime}}g\left(q_{0},\boldsymbol{q}\right)_\mathrm{pNRQCD}g\left(q_{0}^{\prime},\boldsymbol{q}\right)_\mathrm{pNRQCD}\\
 &  & \times\left\{ \tilde{U}_{s}\left(t-v\right)r_{i}\tilde{U}_{o}^{\dagger}\left(t-v\right)\tilde{U}_{o}\left(t-v^{\prime}\right)r_{i}^{\prime}\tilde{U}_{s}^{\dagger}\left(t-v^{\prime}\right),\rho_{s}\left(t\right)\right\} \\
\\
\end{array}\label{eq:192},
\end{equation} 
where $\boldsymbol{s}$ in the NRQCD expression denotes the dipole size, in analogy with $\boldsymbol{r}$ in the pNRQCD expressions; see Ref.~\cite{daddi2024}.

Comparing Eqs.~(\ref{eq:191}) and (\ref{eq:192}), we see that the two expressions become identical provided that
\begin{equation}
\delta_{ij} \int_{\boldsymbol{q}}g\left(q_0,\boldsymbol{q}\right)_\mathrm{pNRQCD}g\left(q_0^{\prime},\boldsymbol{q}\right)_\mathrm{pNRQCD}=\int_{\boldsymbol{q}}q_{i}q_{j}g\left(q_0,\boldsymbol{q}\right)_\mathrm{NRQCD}g\left(q_0^{\prime},\boldsymbol{q}\right)_\mathrm{NRQCD}\label{eq:193}, 
\end{equation}
where the subscripts ``NRQCD'' and ``pNRQCD'' denote the jump correlation functions associated with the two effective field theories. Using this relation, the anticommutator in the pNRQCD expression (\ref{eq:192}) becomes
\begin{equation}
\begin{array}{ccl}
\\
& \int_{\boldsymbol{q}}g\left(q_{0},\boldsymbol{q}\right)_\mathrm{pNRQCD}g\left(q_{0}^{\prime},\boldsymbol{q}\right)_\mathrm{pNRQCD}\left\{ \tilde{U}_{s}\left(t-v\right)r_{i}\tilde{U}_{o}^{\dagger}\left(t-v\right)\tilde{U}_{o}\left(t-v^{\prime}\right)r_{i}^{\prime}\tilde{U}_{s}^{\dagger}\left(t-v^{\prime}\right),\rho_{s}\left(t\right)\right\} \\
\\
= &\int_{\boldsymbol{q}} g\left(q_{0},\boldsymbol{q}\right)_\mathrm{NRQCD}g\left(q_{0}^{\prime},\boldsymbol{q}\right)_\mathrm{NRQCD}\left\{ \tilde{U}_{s}\left(t-v\right)r_{i}q_{i}\tilde{U}_{o}^{\dagger}\left(t-v\right)\tilde{U}_{o}\left(t-v^{\prime}\right)r_{j}^{\prime}q_{j}\tilde{U}_{s}^{\dagger}\left(t-v^{\prime}\right),\rho_{s}\left(t\right)\right\} \\
\\
= &\int_{\boldsymbol{q}} g\left(q_{0},\boldsymbol{q}\right)_\mathrm{NRQCD}g\left(q_{0}^{\prime},\boldsymbol{q}\right)_\mathrm{NRQCD}\left\{ \tilde{U}_{s}\left(t-v\right)\boldsymbol{r}.\boldsymbol{q}\tilde{U}_{o}^{\dagger}\left(t-v\right)\tilde{U}_{o}\left(t-v^{\prime}\right)\boldsymbol{r}^{\prime}.\boldsymbol{q}\tilde{U}_{s}^{\dagger}\left(t-v^{\prime}\right),\rho_{s}\left(t\right)\right\} \\
\end{array}
\end{equation}
which is identical to the NRQCD expression (\ref{eq:191}).

The relation (\ref{eq:193}) can be established as follows:
\begin{equation}
\begin{split}
\int_{\boldsymbol{q}}q_{i}q_{j}g_\mathrm{NRQCD}\left(q_0,\boldsymbol{q}\right)&g_\mathrm{NRQCD}\left(q_0^{\prime},\boldsymbol{q}\right)\\&= \int_{\boldsymbol{q}}q_{i}q_{j}\sqrt{\Delta_{\rm NRQCD}\left(q_0,\boldsymbol{q}\right)}\sqrt{\Delta_{\rm NRQCD}\left(q^\prime_0,\boldsymbol{q}\right)}\\
    &=\delta_{ij}\int_{\boldsymbol{q}}\frac{\|\boldsymbol{q}\|^2}{3}\sqrt{\Delta_{\rm NRQCD}\left(q_0,\boldsymbol{q}\right)}\sqrt{\Delta_{\rm NRQCD}\left(q^\prime_0,\boldsymbol{q}\right)}\\
    &=\frac{1}{3}\delta_{ij}\int_{\boldsymbol{q}}  \sqrt{{\|\boldsymbol{q}\|^2}\Delta_{\rm NRQCD}\left(q_0,\boldsymbol{q}\right)}\sqrt{{\|\boldsymbol{q}\|^2}\Delta_{\rm NRQCD}\left(q^\prime_0,\boldsymbol{q}\right)}\\
    &=\frac{1}{3}\delta_{ij}\int_{\boldsymbol{q}}  \sqrt{\Delta_{\rm pNRQCD}\left(q_0,\boldsymbol{q}\right)}\sqrt{\Delta_{\rm pNRQCD}\left(q^\prime_0,\boldsymbol{q}\right)}\\
    &=\delta_{ij}\int_{\boldsymbol{q}} g_\mathrm{pNRQCD}\left(q_0,\boldsymbol{q}\right) g_\mathrm{pNRQCD}\left(q_0^{\prime},\boldsymbol{q}\right),
    \label{proof-Pol}
    \end{split},
\end{equation}
where we have used Eq.~(\ref{eq:10}), which relates the jump correlation function to the QGP correlator in Fourier space. In Eq.~(\ref{proof-Pol}), $\Delta\left(q^\prime_0,\boldsymbol{q}\right)_{\rm NRQCD}\equiv\Delta^>\left(q^\prime_0,\boldsymbol{q}\right)$ is defined in Eqs.~(\ref{convolution-ULE}) and (\ref{thermal-propagators}). Similarly, $\Delta\left(q^\prime_0,\boldsymbol{q}\right)_{\rm pNRQCD}$ denotes the Fourier transform of $\delta^{ab}\Delta\left(t-t^\prime,\boldsymbol{R}-\boldsymbol{R}^\prime\right)_{\rm pNRQCD}\equiv Tr_\mathrm{QGP}\left(\rho_\mathrm{QGP}\tilde{E}_{i}^{a}\left(t,\boldsymbol{R}\right)\tilde{E}_{i}^{b}\left(t^{\prime},\boldsymbol{R}^{\prime}\right)\right)$, while the pNRQCD jump correlation functions are defined in Eq.~(\ref{eq:177}). The second equality follows from the isotropy of the medium.

Applying the same argument to the remaining contributions to the singlet ULE, namely $\mathcal{L}_{so}\left(t\right)\rho_{o}\left(t\right)$ and $\Lambda_{s}\left(t\right)$, shows that the pNRQCD singlet ULE coincides with the small-dipole limit of the corresponding NRQCD ULE.

\subsection{NRQCD vs. pNRQCD octet ULE}

We next compare the octet master equations obtained within the two effective field theories. Applying the same procedure to the NRQCD octet--singlet contribution $\mathcal{L}_{os}\left(t\right)\rho_s\left(t\right)$, given by Eq.~(\ref{eq:126}), shows that it coincides with the corresponding pNRQCD expression, Eq.~(\ref{eq:206}), at leading order in the dipole size.

Similarly, the octet--octet contribution $\mathcal{L}_{oo}\left(t\right)\rho_{o}\left(t\right)$ has the same expression in the two theories. A distinguishing feature of this contribution is the presence of terms proportional to $C_{\boldsymbol{q}.\hat{\boldsymbol{s}}}$ in the NRQCD expression (\ref{eq:127}), which are given by
\begin{equation}
\begin{array}{ccl}
\tilde{\mathcal{L}}^{\mathrm{NRQCD}}_{oo}\left(t\right)\tilde{\rho}_{o}\left(t\right) & \propto &  \frac{N_{c}}{4}\int_{vv^{\prime}q_{0}q_{0}^{\prime}\boldsymbol{q}}e^{i\left(q_{0}-q_{0}^{\prime}\right)t}e^{-iq_{0}v+iq_{0}^{\prime}v^{\prime}}g\left(q_{0},\boldsymbol{q}\right)g\left(q_{0}^{\prime},\boldsymbol{q}\right)\\
  &  & \times\tilde{U}_{o}\left(t-v^{\prime}\right)C_{\boldsymbol{q}.\hat{\boldsymbol{s}}^{\prime}}\tilde{U}_{o}^{\dagger}\left(t-v^{\prime}\right)\tilde{\rho}_{o}\left(t\right)\tilde{U}_{o}\left(t-v\right)C_{\boldsymbol{q}.\hat{\boldsymbol{s}}}\tilde{U}_{o}^{\dagger}\left(t-v\right)\\
\\
  &  & -\frac{N_{c}}{8}\int_{vv^{\prime}q_{0}q_{0}^{\prime}\boldsymbol{q}}e^{i\left(q_{0}-q_{0}^{\prime}\right)t}e^{-iq_{0}v+iq_{0}^{\prime}v^{\prime}}g\left(q_{0},\boldsymbol{q}\right)g\left(q_{0}^{\prime},\boldsymbol{q}\right)\\
  &  & \times\left\{ \tilde{U}_{o}\left(t-v\right)C_{\boldsymbol{q}.\hat{\boldsymbol{s}}}\tilde{U}_{o}^{\dagger}\left(t-v\right)\tilde{U}_{o}\left(t-v^{\prime}\right)C_{\boldsymbol{q}.\hat{\boldsymbol{s}}^{\prime}}\tilde{U}_{o}^{\dagger}\left(t-v^{\prime}\right),\tilde{\rho}_{o}\left(t\right)\right\} \\
\end{array}\label{Eq:4.7}
\end{equation}

In the small-dipole limit, using $C_{\boldsymbol{q}.\hat{\boldsymbol{s}}}\approx2\hat{\boldsymbol{I}}$ and the unitarity of the evolution operator, $\hat{U}^\dagger\hat{U}=\hat{I}$, the two terms in Eq.~(\ref{Eq:4.7}) cancel exactly.

The remaining terms in the NRQCD octet--octet contribution (\ref{eq:127}), which are proportional to $S_{\boldsymbol{q}.\hat{\boldsymbol{s}}}$, reduce in the small-dipole limit, together with Eq.~(\ref{eq:193}), to the same expression as the pNRQCD octet--octet contribution (\ref{pNRQCD-octet-octet-element}).

Finally, we compare the octet Lamb-shift contribution $\Lambda_{o}\left(t\right)$ in the two theories. In the small-dipole limit, a partial difference remains because of the term proportional to $C_{\boldsymbol{q}.\hat{\boldsymbol{s}}}$ in the NRQCD expression (\ref{eq:125}). However, this difference does not affect the octet master equation because the resulting contribution commutes with the density matrix.

Indeed, taking the small-dipole limit of the NRQCD octet Lamb-shift contribution (\ref{eq:125}), we find that it differs from the pNRQCD expression (\ref{pNRQCD-octet-LS}) by
\begin{equation}
\frac{N_{c}}{8i}\int_{vv^{\prime}q_{0}q_{0}^{\prime}\boldsymbol{q}}\text{sign}\left(v-v^{\prime}\right)e^{i\left(q_{0}-q_{0}^{\prime}\right)t}e^{-iq_{0}v+iq_{0}^{\prime}v^{\prime}}g\left(q_{0},\boldsymbol{q}\right)g\left(q_{0}^{\prime},\boldsymbol{q}\right)\hat{I},
\end{equation}
where we have again used $C_{\boldsymbol{q}.\hat{\boldsymbol{s}}}\approx2\hat{\boldsymbol{I}}$ and $\hat{U}^\dagger\hat{U}=\hat{I}$. Since this contribution is proportional to the identity operator, it commutes with the density matrix and therefore drops out of the commutator in Eq.~(\ref{eq:octet-equation-com-integrated-out}). Thus, the NRQCD octet Lamb-shift contribution (\ref{eq:125}) coincides with the pNRQCD expression (\ref{pNRQCD-octet-LS}) in the small-dipole limit as far as the dynamics is concerned.

Based on the comparison of the various contributions to the octet ULE, we conclude that the pNRQCD octet ULE coincides with the small-dipole limit of the NRQCD octet ULE.

Therefore, we have shown that the singlet--octet pNRQCD ULEs are recovered as the small-dipole limit of the corresponding NRQCD ULEs. Consequently, a numerical solution of the latter provides a more general description of the quarkonium dynamics, as its validity extends beyond the small-dipole approximation employed in the pNRQCD master equation.\footnote{Nevertheless, an important advantage of the pNRQCD ULEs over the NRQCD ones is their manifest gauge invariance, whereas the gauge invariance of the NRQCD formulation remains ambiguous.}

\section{Boltzmann equations}
\label{Semiclassical-limit-of-NRQCD-ULEs}

In this section, we derive the semiclassical limit of the NRQCD universal Lindblad equations, thereby obtaining the corresponding coupled singlet--octet Boltzmann equations. To our knowledge, this is the first derivation of a semiclassical transport description directly from the universal Lindblad framework. Since no small-dipole approximation is invoked, the resulting equations extend beyond that limit. We then consider their small-dipole limit and compare the resulting expressions with the Boltzmann equations derived by Yao \textit{et al.}~\cite{Yao:2018nmy,Yao:2021lus,akamatsu2022quarkonium}.

\subsection{From NRQCD ULEs to Boltzmann equations}

Before implementing the semiclassical approximation, we first project the singlet--octet ULEs onto the center-of-mass and energy bases. The resulting equations can be written as follows:
\begin{equation}
\begin{split}
\frac{d\langle n,\boldsymbol{P}_1|{\rho}_{s}\left(t\right)|n^{\prime},\boldsymbol{P}_2\rangle }{dt}=&-i\langle n,\boldsymbol{P}_1|\left[{H}_{s}+{\Lambda}_{s},{\rho_{s}}\left(t\right)\right]|n^{\prime},\boldsymbol{P}_2\rangle  +\langle n,\boldsymbol{P}_1|{\mathcal{L}}_{ss}\left(t\right){\rho}_{s}\left(t\right)|n^{\prime},\boldsymbol{P}_2\rangle  \\
&+\langle n,\boldsymbol{P}_1|\mathcal{{L}}_{so}\left(t\right){\rho}_{o}\left(t\right)|n^{\prime},\boldsymbol{P}_2\rangle 
\end{split}
\label{singlet-eq-projected-on-energy},
\end{equation}
\begin{equation}
\begin{split}
\frac{d\langle \boldsymbol{k},\boldsymbol{P}_1|{\rho}_{o}\left(t\right)|\boldsymbol{k}^{\prime},\boldsymbol{P}_2\rangle }{dt}=&-i\langle \boldsymbol{k},\boldsymbol{P}_1|\left[{H}_{o}+{\Lambda}_{o},{\rho}_{o}\left(t\right)\right]|\boldsymbol{k}^{\prime},\boldsymbol{P}_2\rangle +\langle \boldsymbol{k},\boldsymbol{P}_1|{\mathcal{L}}_{os}\left(t\right){\rho}_{s}\left(t\right)|\boldsymbol{k}^{\prime},\boldsymbol{P}_2\rangle \\
&+\langle \boldsymbol{k},\boldsymbol{P}_1|\mathcal{{L}}_{oo}\left(t\right){\rho}_{o}\left(t\right)|\boldsymbol{k}^{\prime},\boldsymbol{P}_2\rangle \label{octet-eq-projected-on-energy}
\end{split},
\end{equation}
where $|n\rangle$ and $|\boldsymbol{k}\rangle$ denote the eigenstates of the singlet and octet sectors, respectively, while $\boldsymbol{P}_i$ denotes the center-of-mass momentum. The details of the projection onto the center-of-mass and energy bases can be found in Refs.~\cite{daddi2024,daddigeneral}, where the techniques developed in Ref.~\cite{blaizot2018approach} were closely followed.\footnote{It is important to note that, in deriving Eqs.~(\ref{singlet-eq-projected-on-energy}) and (\ref{octet-eq-projected-on-energy}), we project onto the center-of-mass basis without integrating out this degree of freedom, in contrast to the procedure followed in Refs.~\cite{daddi2024,daddigeneral}.}

The expressions for the singlet Lamb-shift and transition terms in Eq.~(\ref{singlet-eq-projected-on-energy}) are given by
\begin{equation}
\begin{array}{ccl}
\langle n,\boldsymbol{P}_1|\left[{\Lambda}_{s},{\rho_{s}}\left(t\right)\right]|n^{\prime},\boldsymbol{P}_2\rangle  & = & -C_{F}\underset{m}{\sum}\int_{\boldsymbol{q}\boldsymbol{p}}\text{P}\int_{q_{0}^{\prime}}\frac{g\left(q_{0}^{\prime}+E_{n}-E_{m},\boldsymbol{q}\right)g\left(q_{0}^{\prime},\boldsymbol{q}\right)}{q_{0}^{\prime}-E_{m}+\frac{k^{2}}{M}}\\
 &  & \times\langle n\left|S_{\boldsymbol{q}.\hat{\boldsymbol{s}}}\left|\boldsymbol{p}\rangle \langle \boldsymbol{p}\right|S_{\boldsymbol{q}.\hat{\boldsymbol{s}}^{\prime}}\left|m\rangle \langle m,\boldsymbol{P}_1\right|{\rho}_{s}\left(t\right)\right|n^{\prime},\boldsymbol{P}_2\rangle \\
\\
  &  & +C_{F}\underset{m}{\sum}\int_{\boldsymbol{q}\boldsymbol{p}}\text{P}\int_{q_{0}^{\prime}}\frac{g\left(q_{0}^{\prime}+E_{m}-E_{n^{\prime}},\boldsymbol{q}\right)g\left(q_{0}^{\prime},\boldsymbol{q}\right)}{q_{0}^{\prime}-E_{n^{\prime}}+\frac{k^{2}}{M}}\\
 &  & \times\langle n,\boldsymbol{P}_1|{\rho}_{s}\left(t\right)|m,\boldsymbol{P}_2\rangle \langle m|S_{\boldsymbol{q}.\hat{\boldsymbol{s}}}|\boldsymbol{p}\rangle \langle \boldsymbol{p}|S_{\boldsymbol{q}.\hat{\boldsymbol{s}}^{\prime}}|n^{\prime}\rangle
\end{array},\label{singlet-LS-on-energy-basis}
\end{equation}
\begin{equation}
\begin{array}{ccl}
\langle n,\boldsymbol{P}_1|{\mathcal{L}}_{ss}\left(t\right){\rho_{s}}\left(t\right)|n^{\prime},\boldsymbol{P}_2\rangle  & = & \frac{-C_{F}}{2}\underset{m}{\sum}\int_{\boldsymbol{q}\boldsymbol{p}}g\left(E_{n}-\frac{p^{2}}{M},\boldsymbol{q}\right)g\left(E_{m}-\frac{p^{2}}{M},\boldsymbol{q}\right)\\
 &  & \times\langle n|S_{\boldsymbol{q}.\hat{\boldsymbol{s}}}|\boldsymbol{p}\rangle \langle \boldsymbol{p}|S_{\boldsymbol{q}.\hat{\boldsymbol{s}}^{\prime}}|m\rangle \langle m,\boldsymbol{P}_1|{\rho}_{s}\left(t\right)|n^{\prime},\boldsymbol{P}_2\rangle \\
\\
  &  & \frac{-C_{F}}{2}\underset{m}{\sum}\int_{\boldsymbol{q}\boldsymbol{p}}g\left(E_{m}-\frac{p^{2}}{M},\boldsymbol{q}\right)g\left(E_{n^{\prime}}-\frac{p^{2}}{M},\boldsymbol{q}\right)\\
 &  & \times\langle n,\boldsymbol{P}_1|{\rho}_{s}\left(t\right)|m,\boldsymbol{P}_2\rangle \langle m|S_{\boldsymbol{q}.\hat{\boldsymbol{s}}}|\boldsymbol{p}\rangle \langle \boldsymbol{p}|S_{\boldsymbol{q}.\hat{\boldsymbol{s}}^{\prime}}|n^{\prime}\rangle 
\end{array},\label{193}
\end{equation}
\begin{equation}
\begin{array}{ccl}
\langle n,\boldsymbol{P}_1|{\mathcal{L}}_{so}\left(t\right){\rho_{o}}\left(t\right)|n^{\prime},\boldsymbol{P}_2\rangle  & = & C_{F}\int_{\boldsymbol{q}\boldsymbol{p}\boldsymbol{p}^{\prime}} g\left(\frac{p^{2}}{M}-E_{n^{\prime}},\boldsymbol{q}\right)g\left(\frac{p^{\prime2}}{M}-E_{n},\boldsymbol{q}\right)\\
 &  & \times\langle n|S_{\boldsymbol{q}.\hat{\boldsymbol{s}}^{\prime}}|\boldsymbol{p}^{\prime}\rangle \langle \boldsymbol{p}^{\prime},\boldsymbol{P}_1|{\rho}_{o}\left(t\right)|\boldsymbol{p},\boldsymbol{P}_2\rangle \langle \boldsymbol{p}|S_{\boldsymbol{q}.\hat{\boldsymbol{s}}}|n^{\prime}\rangle\\ \label{194}
\end{array}.
\end{equation}

The expressions for the various terms in the octet ULE in Eq.~(\ref{octet-eq-projected-on-energy}), which are rather lengthy, are collected in Appendix~\ref{Appendix-The octet ULE matrix elements}.

The singlet--octet Boltzmann equations are obtained by performing a Wigner transformation of Eqs.~(\ref{singlet-eq-projected-on-energy}) and (\ref{octet-eq-projected-on-energy}), followed by the semiclassical approximation, following the procedure developed in Refs.~\cite{yao2021semiclassical-transport,Yao:2021lus,akamatsu2022quarkonium}.

We first define the Wigner transforms of the singlet and octet density matrices. In the semiclassical limit, we take the singlet density matrix to be diagonal in the energy eigenbasis, $\rho_s\propto |n\rangle\langle n|$, and neglect the off-diagonal coherence terms. Accordingly, the Wigner transforms of the singlet and octet density matrices are given by
\begin{equation}
f_n\left(t, \boldsymbol{R},\boldsymbol{P}\right)=\int_{\Delta\boldsymbol{P}}e^{i\Delta \boldsymbol{P}.\boldsymbol{R}}\langle\boldsymbol{P}_1,n|\rho_s\left(t\right)|\boldsymbol{P}_2,n\rangle \label{wigner-singlet}
\end{equation}
\begin{equation}
f_o\left(t, \boldsymbol{R},\boldsymbol{P},\boldsymbol{r},\tilde{\boldsymbol{p}}\right)=\int_{\Delta\boldsymbol{P}\Delta\boldsymbol{p}}e^{i\Delta \boldsymbol{P}.\boldsymbol{R}+i\Delta \boldsymbol{p}.\boldsymbol{r}}\langle\boldsymbol{P}_1,\boldsymbol{p}|\rho_o\left(t\right)|\boldsymbol{P}_2,\boldsymbol{p}^\prime\rangle,\label{wigner-octet}
\end{equation}
where $(\boldsymbol{R},\boldsymbol{P})$ and $(\boldsymbol{r},\tilde{\boldsymbol{p}})$ are the center-of-mass and relative coordinates, respectively, with $\boldsymbol{P}=\frac{\boldsymbol{P}_1+\boldsymbol{P}_2}{2}$, $\tilde{\boldsymbol{p}}=\frac{\boldsymbol{p}+\boldsymbol{p}^\prime}{2}$, $\Delta\boldsymbol{P}=\boldsymbol{P}_2-\boldsymbol{P}_1$, and $\Delta\boldsymbol{p}=\boldsymbol{p}^\prime-\boldsymbol{p}$.

Equivalently, the Wigner transforms can be written as
\begin{equation}
f_n\left(t, \boldsymbol{R},\boldsymbol{P}\right)=\int_{\Delta\boldsymbol{P}}e^{i\Delta \boldsymbol{P}.\boldsymbol{R}}\langle\boldsymbol{P},n|\rho_s\left(t\right)|\boldsymbol{P}+\Delta \boldsymbol{P},n\rangle ,
\end{equation}
and, similarly, for the octet channel,
\begin{equation}
f_o\left(t, \boldsymbol{R},\boldsymbol{P},\boldsymbol{r},\tilde{\boldsymbol{p}}\right)=\int_{\Delta\boldsymbol{P}\Delta\boldsymbol{p}}e^{i\Delta \boldsymbol{P}.\boldsymbol{R}+i\Delta \boldsymbol{p}.\boldsymbol{r}}\langle\boldsymbol{P},\boldsymbol{p}|\rho_o\left(t\right)|\boldsymbol{P}+\Delta \boldsymbol{P},\tilde{\boldsymbol{p}}+\Delta \boldsymbol{p}\rangle,
\end{equation}
These equivalent definitions will be useful in the derivation of the Boltzmann equations.

We first outline the main steps and approximations involved in deriving the singlet Boltzmann equation from Eq.~(\ref{singlet-eq-projected-on-energy}). In addition to the assumption that the density matrix is diagonal in the energy basis, the semiclassical limit requires setting $\boldsymbol{s}=\boldsymbol{s}^\prime$ in Eqs.~(\ref{singlet-eq-projected-on-energy}) and (\ref{octet-eq-projected-on-energy}), since the coherence length, defined by $\boldsymbol{y}=\boldsymbol{s}^\prime-\boldsymbol{s}$, vanishes in this limit.\footnote{This follows from the fact that the classical limit involves only the diagonal matrix elements appearing in Eq.~(A1) of Appendix A of Ref.~\cite{blaizot2018approach}, which requires $\boldsymbol{s}=\boldsymbol{s}^\prime$; see Eq.~(A2) therein.}

Consequently, the Wigner transform of the singlet ULE in the semiclassical limit yields the following expression for the singlet--singlet component:
\begin{equation}
\begin{split}
\int_{\Delta\boldsymbol{P}}e^{i\Delta \boldsymbol{P}.\boldsymbol{R}}\langle\boldsymbol{P},n|\mathcal{L}_{ss}\rho_s\left(t\right)|\boldsymbol{P}+\Delta \boldsymbol{P},n\rangle=& -C_{F}\int_{\boldsymbol{q}\boldsymbol{p}}\left[g\left(E_{n}-\frac{p^{2}}{M},\boldsymbol{q}\right)\right]^2\\
& \times|\langle n|S_{\boldsymbol{q}.\hat{\boldsymbol{s}}}|\boldsymbol{p}\rangle|^2f_n\left(t, \boldsymbol{R},\boldsymbol{P}\right) 
\end{split},
\end{equation}
with $|\langle n|S_{\boldsymbol{q}.\hat{\boldsymbol{s}}}|\boldsymbol{p}\rangle|^2=\langle n|S_{\boldsymbol{q}.\hat{\boldsymbol{s}}}|\boldsymbol{p}\rangle \langle \boldsymbol{p}|S_{\boldsymbol{q}.\hat{\boldsymbol{s}}}|n\rangle$.

The octet--singlet contribution is given by
\begin{equation}
\begin{split}
\int_{\Delta\boldsymbol{P}}e^{i\Delta \boldsymbol{P}.\boldsymbol{R}}\langle\boldsymbol{P},n|\mathcal{L}_{so}\rho_o\left(t\right)|\boldsymbol{P}+\Delta \boldsymbol{P},n\rangle=& C_{F}\int_{\boldsymbol{q}\boldsymbol{p}}\left[g\left(\frac{p^{2}}{M}-E_{n},\boldsymbol{q}\right)\right]^2\\
& \times|\langle n|S_{\boldsymbol{q}.\hat{\boldsymbol{s}}}|\boldsymbol{p}\rangle|^2f_o\left(t, \boldsymbol{R},\boldsymbol{P},\boldsymbol{0},\boldsymbol{p}\right)
\end{split}.\label{5.11}
\end{equation}
Here, following Ref.~\cite{akamatsu2022quarkonium}, we treat the unbound octet state as an extended state whose wave function is localized in momentum space, i.e. $\Delta\boldsymbol{p}\simeq0$.\footnote{In this case, $\tilde{\boldsymbol{p}}\simeq\boldsymbol{p}$.}

Alternatively, the semiclassical approximation can be implemented directly at the level of Eq.~(\ref{194}) by performing the following semiclassical, or gradient, expansion of the octet Wigner function~\cite{haug2008quantum,Yao:2018nmy}:
\begin{equation}
f_o\left(t, \boldsymbol{R},\boldsymbol{P},\boldsymbol{r},\tilde{\boldsymbol{p}}\right)=f_o\left(t, \boldsymbol{R},\boldsymbol{P},\boldsymbol{r}_0,\tilde{\boldsymbol{p}}\right)+\left(\boldsymbol{r}-\boldsymbol{r}_0\right).\nabla_{\boldsymbol{r}_0}f_o\left(t, \boldsymbol{R},\boldsymbol{P},\boldsymbol{r}_0,\tilde{\boldsymbol{p}}\right)+\cdots,\label{gradient-expansion-main} 
\end{equation}
where Eq.~(\ref{5.11}) is recovered by retaining only the leading-order term in this expansion. The details of this derivation are given in Appendix~\ref{Appendix-computation-details-semiclassical-limit}. This is also the procedure adopted in Refs.~\cite{Yao:2018nmy,yao2021semiclassical-transport}.

Retaining higher-order terms in Eq.~(\ref{gradient-expansion-main}) provides a systematic way of incorporating quantum corrections to the semiclassical dynamics. The leading quantum correction is derived in Sec.~\ref{section-quantum-corrections}.

In the semiclassical limit, the unitary evolution contributes only through the center-of-mass kinetic term, while the Lamb-shift contribution vanishes. The singlet Boltzmann equation therefore takes the form
\begin{equation}
\begin{split}
\left[\frac{\partial}{\partial t}+\frac{\boldsymbol{P}}{2M}.\nabla_{\boldsymbol{R}}\right]f_n\left(t, \boldsymbol{R},\boldsymbol{P}\right)=&-C_{F}\int_{\boldsymbol{q}\boldsymbol{p}}\left[g\left(E_{n}-\frac{p^{2}}{M},\boldsymbol{q}\right)\right]^2|\langle n|S_{\boldsymbol{q}.\hat{\boldsymbol{s}}}|\boldsymbol{p}\rangle|^2f_n\left(t, \boldsymbol{R},\boldsymbol{P}\right)\\&+C_{F}\int_{\boldsymbol{q}\boldsymbol{p}}\left[g\left(\frac{p^{2}}{M}-E_{n},\boldsymbol{q}\right)\right]^2|\langle n|S_{\boldsymbol{q}.\hat{\boldsymbol{s}}}|\boldsymbol{p}\rangle|^2
f_o\left(t, \boldsymbol{R},\boldsymbol{P},\boldsymbol{0},\boldsymbol{p}\right) 
\end{split} 
\end{equation}

We now derive the octet Boltzmann equation from Eq.~(\ref{octet-eq-projected-on-energy}); see Appendix~\ref{Appendix-The octet ULE matrix elements}. The Wigner transform of the singlet--octet matrix element yields
\begin{equation}
\begin{split}
\int_{\Delta\boldsymbol{P}\Delta\boldsymbol{k}}e^{i\Delta \boldsymbol{P}.\boldsymbol{R}+i\Delta \boldsymbol{k}.\boldsymbol{r}}&\langle\boldsymbol{P},\boldsymbol{k}|\mathcal{L}_{os}\rho_s\left(t\right)|\boldsymbol{P}+\Delta \boldsymbol{P},\boldsymbol{k}+\Delta\boldsymbol{k}\rangle= \frac{1}{2N_C}\delta\left(\boldsymbol{r}\right)\sum_{n}\int_{\boldsymbol{q}}\left[g\left(E_{n}-\frac{k^{2}}{M},\boldsymbol{q}\right)\right]^2\\
& \times|\langle n|S_{\boldsymbol{q}.\hat{\boldsymbol{s}}}|\boldsymbol{k}\rangle|^2f_n\left(t, \boldsymbol{R},\boldsymbol{P}\right) 
\end{split},  
\end{equation}
where the delta function $\delta(\boldsymbol{r})$ arises from the integral $\int_{\Delta\boldsymbol{k}}e^{i\Delta\boldsymbol{k}\cdot\boldsymbol{r}}$ after approximating $\boldsymbol{k}\simeq\boldsymbol{k}^\prime$ in the jump-correlation functions appearing in Eq.~(\ref{l_os-rho_s-element}). This approximation follows from treating the unbound octet state as an extended state, as discussed below Eq.~(\ref{5.11}); see Ref.~\cite{akamatsu2022quarkonium} for further details. Applying the same approximation to the octet--octet matrix element yields its semiclassical limit and, consequently, the octet Boltzmann equation
\begin{equation}
\begin{split}
\left[\frac{\partial}{\partial t}+\frac{\boldsymbol{P}}{2M}.\nabla_{\boldsymbol{R}}+2\frac{\boldsymbol{k}}{M}.\nabla_{\boldsymbol{r}}\right]&f_o\left(t, \boldsymbol{R},\boldsymbol{P},\boldsymbol{r},\boldsymbol{k}\right) \\&=\frac{1}{2N_C}\delta\left(\boldsymbol{r}\right)\sum_{n}\int_{\boldsymbol{q}}\left[g\left(E_{n}-\frac{k^{2}}{M},\boldsymbol{q}\right)\right]^2|\langle n|S_{\boldsymbol{q}.\hat{\boldsymbol{s}}}|\boldsymbol{k}\rangle|^2f_n\left(t, \boldsymbol{R},\boldsymbol{P}\right) \\&+\frac{N_c^2-4}{4N_c}\delta\left(\boldsymbol{r}\right)\int_{\boldsymbol{q}\boldsymbol{p}}\left[g\left(\frac{p^{2}}{M}-\frac{k^{2}}{M},\boldsymbol{q}\right)\right]^2|\langle \boldsymbol{k}|S_{\boldsymbol{q}.\hat{\boldsymbol{s}}}|\boldsymbol{p}\rangle|^2
f_o\left(t, \boldsymbol{R},\boldsymbol{P},\boldsymbol{0},\boldsymbol{k}\right)\\&-\frac{N_c^2-4}{4N_c}\delta\left(\boldsymbol{r}\right)\int_{\boldsymbol{q}\boldsymbol{p}}\left[g\left(\frac{k^{2}}{M}-\frac{p^{2}}{M},\boldsymbol{q}\right)\right]^2|\langle \boldsymbol{k}|S_{\boldsymbol{q}.\hat{\boldsymbol{s}}}|\boldsymbol{p}\rangle|^2
f_o\left(t, \boldsymbol{R},\boldsymbol{P},\boldsymbol{0},\boldsymbol{k}\right)\\&-\frac{1}{2N_c}\delta\left(\boldsymbol{r}\right)\sum_n\int_{\boldsymbol{q}}\left[g\left(\frac{k^{2}}{M}-E_n,\boldsymbol{q}\right)\right]^2|\langle n|S_{\boldsymbol{q}.\hat{\boldsymbol{s}}}|\boldsymbol{k}\rangle|^2
f_o\left(t, \boldsymbol{R},\boldsymbol{P},\boldsymbol{0},\boldsymbol{k}\right)\\&+\frac{N_c }{4}\delta\left(\boldsymbol{r}\right)\int_{\boldsymbol{q}\boldsymbol{p}}\left[g\left(\frac{p^{2}}{M}-\frac{k^{2}}{M},\boldsymbol{q}\right)\right]^2|\langle \boldsymbol{k}|C_{\boldsymbol{q}.\hat{\boldsymbol{s}}}|\boldsymbol{p}\rangle|^2
f_o\left(t, \boldsymbol{R},\boldsymbol{P},\boldsymbol{0},\boldsymbol{k}\right) \\&-\frac{N_c}{4}\delta\left(\boldsymbol{r}\right)\int_{\boldsymbol{q}\boldsymbol{p}}\left[g\left(\frac{k^{2}}{M}-\frac{p^{2}}{M},\boldsymbol{q}\right)\right]^2|\langle \boldsymbol{k}|C_{\boldsymbol{q}.\hat{\boldsymbol{s}}}|\boldsymbol{p}\rangle|^2
f_o\left(t, \boldsymbol{R},\boldsymbol{P},\boldsymbol{0},\boldsymbol{k}\right)
\end{split}.
\end{equation}

The above equations can be cast into a more compact form using Eq.~(\ref{eq:10}) together with the Kubo--Martin--Schwinger (KMS) relation
\begin{equation}
\Delta^{>}\left(q_{0},\boldsymbol{q}\right)=e^{\beta q_{0}}\Delta^{<}\left(q_{0},\boldsymbol{q}\right),
\label{KMS}
\end{equation}
which implies
\[
\Delta^{>}\left(-q_{0},\boldsymbol{q}\right)
=
\Delta^{<}\left(q_{0},\boldsymbol{q}\right)
=
e^{-\beta q_{0}}\Delta^{>}\left(q_{0},\boldsymbol{q}\right),
\]
with $\beta=1/T$.

Using these relations, the singlet--octet universal Boltzmann equations take the form
\begin{equation}
\begin{split}
\left[\frac{\partial}{\partial t}+\frac{\boldsymbol{P}}{2M}.\nabla_{\boldsymbol{R}}\right]f_n\left(t, \boldsymbol{R},\boldsymbol{P}\right)=&-{C_{F}}\int_{\boldsymbol{q}\boldsymbol{p}}\Delta^>\left(E_{n}-\frac{p^{2}}{M},\boldsymbol{q}\right)|\langle n|S_{\boldsymbol{q}.\hat{\boldsymbol{s}}}|\boldsymbol{p}\rangle|^2\\
& \times \left(f_n\left(t, \boldsymbol{R},\boldsymbol{P}\right)-e^{-\beta\left(E_{n}-\frac{p^{2}}{M}\right) }f_o\left(t, \boldsymbol{R},\boldsymbol{P},\boldsymbol{0},\boldsymbol{p}\right) \right)
\end{split} \label{boltz-singlet-no-recoil}
\end{equation}

\begin{equation}
\begin{split}
\left[\frac{\partial}{\partial t}+\frac{\boldsymbol{P}}{2M}.\nabla_{\boldsymbol{R}}+2\frac{\boldsymbol{k}}{M}.\nabla_{\boldsymbol{r}}\right]&f_o\left(t, \boldsymbol{R},\boldsymbol{P},\boldsymbol{r},\boldsymbol{k}\right) \\&=\frac{1}{2N_C}\delta\left(\boldsymbol{r}\right)\sum_{n}\int_{\boldsymbol{q}}\Delta^>\left(E_{n}-\frac{k^{2}}{M},\boldsymbol{q}\right)|\langle n|S_{\boldsymbol{q}.\hat{\boldsymbol{s}}}|\boldsymbol{k}\rangle|^2
\\& \times \left(f_n\left(t, \boldsymbol{R},\boldsymbol{P}\right)-e^{-\beta\left(E_{n}-\frac{k^{2}}{M}\right) }f_o\left(t, \boldsymbol{R},\boldsymbol{P},\boldsymbol{0},\boldsymbol{k}\right)\right)
\\&+\frac{N_c^2-4}{4 N_c}\delta\left(\boldsymbol{r}\right)\int_{\boldsymbol{q}\boldsymbol{p}}\Delta^>\left(\frac{p^{2}}{M}-\frac{k^{2}}{M},\boldsymbol{q}\right)|\langle \boldsymbol{k}|S_{\boldsymbol{q}.\hat{\boldsymbol{s}}}|\boldsymbol{p}\rangle|^2\\
& \times \left(1-e^{-\beta\left(\frac{p^{2}}{M}-\frac{k^{2}}{M}\right) }\right)
f_o\left(t, \boldsymbol{R},\boldsymbol{P},\boldsymbol{0},\boldsymbol{k}\right)\\&+\frac{N_c }{4}\delta\left(\boldsymbol{r}\right)\int_{\boldsymbol{q}\boldsymbol{p}}\Delta^>\left(\frac{p^{2}}{M}-\frac{k^{2}}{M},\boldsymbol{q}\right)|\langle \boldsymbol{k}|C_{\boldsymbol{q}.\hat{\boldsymbol{s}}}|\boldsymbol{p}\rangle|^2\\
& \times \left(1-e^{-\beta\left(\frac{p^{2}}{M}-\frac{k^{2}}{M}\right) }\right)
f_o\left(t, \boldsymbol{R},\boldsymbol{P},\boldsymbol{0},\boldsymbol{k}\right)
\end{split}.\label{boltz-octet-no-recoil}
\end{equation}

Equations~(\ref{boltz-singlet-no-recoil}) and (\ref{boltz-octet-no-recoil}) constitute the general semiclassical counterpart of the NRQCD universal Lindblad equations. Since they are derived without invoking the small-dipole approximation, they remain valid beyond that limit. To facilitate comparison with the results of Yao \textit{et al.}~\cite{Yao:2018nmy,yao2021semiclassical-transport}, we now specialize them to the small-dipole regime.

\subsection{The small-dipole limit}

The small-dipole limit is implemented by expanding the functions $S_{\boldsymbol{q}\cdot\hat{\boldsymbol{s}}}$ and $C_{\boldsymbol{q}\cdot\hat{\boldsymbol{s}}}$ according to Eq.~(\ref{eq:small-s-limit}). In this limit, the last term in the octet equation~(\ref{boltz-octet-no-recoil}) vanishes because
\begin{equation}
\langle \boldsymbol{k}|C_{\boldsymbol{q}\cdot\hat{\boldsymbol{s}}}|\boldsymbol{p}\rangle
\simeq
2\langle \boldsymbol{k}|\hat{I}|\boldsymbol{p}\rangle
=
2\delta(\boldsymbol{k}-\boldsymbol{p}).
\end{equation}

The corresponding Boltzmann equations then reduce to
\begin{equation}
\begin{split}
\left[\frac{\partial}{\partial t}+\frac{\boldsymbol{P}}{2M}.\nabla_{\boldsymbol{R}}\right]f_n\left(t, \boldsymbol{R},\boldsymbol{P}\right)=&-{C_{F}}\int_{\boldsymbol{q}\boldsymbol{p}}q_iq_j\Delta^>\left(E_{n}-\frac{p^{2}}{M},\boldsymbol{q}\right)\langle n|\hat{\boldsymbol{s}}_i|\boldsymbol{p}\rangle\langle \boldsymbol{p} |{\hat{\boldsymbol{s}}_j}|n\rangle\\
& \times \left(f_n\left(t, \boldsymbol{R},\boldsymbol{P}\right)-e^{-\beta\left(E_{n}-\frac{p^{2}}{M}\right) }f_o\left(t, \boldsymbol{R},\boldsymbol{P},\boldsymbol{0},\boldsymbol{p}\right) \right)
\end{split}  \label{boltz-singlet-no-recoil-small-dipol},
\end{equation}
\begin{equation}
\begin{split}
\left[\frac{\partial}{\partial t}+\frac{\boldsymbol{P}}{2M}.\nabla_{\boldsymbol{R}}+2\frac{\boldsymbol{k}}{M}.\nabla_{\boldsymbol{r}}\right]&f_o\left(t, \boldsymbol{R},\boldsymbol{P},\boldsymbol{r},\boldsymbol{k}\right) \\&=\frac{1}{2N_C}\delta\left(\boldsymbol{r}\right)\sum_{n}\int_{\boldsymbol{q}}q_iq_j\Delta^>\left(E_{n}-\frac{k^{2}}{M},\boldsymbol{q}\right)\langle n|\hat{\boldsymbol{s}}_i|\boldsymbol{k}\rangle\langle\boldsymbol{k}|\hat{\boldsymbol{s}}_j|n\rangle
\\& \times \left(f_n\left(t, \boldsymbol{R},\boldsymbol{P}\right)-e^{-\beta\left(E_{n}-\frac{k^{2}}{M}\right) }f_o\left(t, \boldsymbol{R},\boldsymbol{P},\boldsymbol{0},\boldsymbol{k}\right)\right)
\\&+\frac{N_c^2-4}{4 N_c}\delta\left(\boldsymbol{r}\right)\int_{\boldsymbol{q}\boldsymbol{p}}q_iq_j\Delta^>\left(\frac{p^{2}}{M}-\frac{k^{2}}{M},\boldsymbol{q}\right)\langle \boldsymbol{k}|\hat{\boldsymbol{s}}_i|\boldsymbol{p}\rangle\langle \boldsymbol{p}|\hat{\boldsymbol{s}}_j|\boldsymbol{k}\rangle\\
& \times \left(1-e^{-\beta\left(\frac{p^{2}}{M}-\frac{k^{2}}{M}\right) }\right)
f_o\left(t, \boldsymbol{R},\boldsymbol{P},\boldsymbol{0},\boldsymbol{k}\right)
\end{split}.\label{boltz-octet-no-recoil-small-dipole}
\end{equation}

To proceed, we use Eq.~(\ref{eq:193}), which relates the pNRQCD and NRQCD jump correlators. Setting $q_0=q^\prime_0$ in this relation and using Eq.~(\ref{eq:10}), we obtain
\begin{equation}
\begin{split}
\int_{\boldsymbol{q}}q_{i}q_{j}\Delta_{\rm NRQCD}\left(q_0,\boldsymbol{q}\right)=\delta_{ij}\int_{\boldsymbol{q}}\frac{\|\boldsymbol{q}\|^2}{3}\Delta_{\rm NRQCD}\left(q_0,\boldsymbol{q}\right)=\delta_{ij}\int_{\boldsymbol{q}}  \Delta_{\rm pNRQCD}\left(q_0,\boldsymbol{q}\right)
\label{proof-Pol-2}
\end{split}.
\end{equation}

Here, $\Delta\left(q^\prime_0,\boldsymbol{q}\right)_{\rm NRQCD}\equiv\Delta^>\left(q^\prime_0,\boldsymbol{q}\right)$ is defined in Eqs.~(\ref{convolution-ULE}) and (\ref{thermal-propagators}). Similarly, $\Delta\left(q^\prime_0,\boldsymbol{q}\right)_{\rm pNRQCD}$ denotes the Fourier transform of $\delta^{ab}\Delta\left(t-t^\prime,\boldsymbol{R}-\boldsymbol{R}^\prime\right)_{\rm pNRQCD}\equiv Tr_\mathrm{QGP}\left(\rho_\mathrm{QGP}\tilde{E}_{i}^{a}\left(t,\boldsymbol{R}\right)\tilde{E}_{i}^{b}\left(t^{\prime},\boldsymbol{R}^{\prime}\right)\right)$. The second equality follows from the isotropy of the medium.

Using Eq.~(\ref{proof-Pol-2}), the Boltzmann equations (\ref{boltz-singlet-no-recoil-small-dipol}) and (\ref{boltz-octet-no-recoil-small-dipole}) become
\begin{equation}
\begin{split}
\left[\frac{\partial}{\partial t}+\frac{\boldsymbol{P}}{2M}.\nabla_{\boldsymbol{R}}\right]f_n\left(t, \boldsymbol{R},\boldsymbol{P}\right)=&-{C_{F}}\delta_{ij}\int_{\boldsymbol{q}\boldsymbol{p}}\Delta_{\rm pNRQCD}\left(E_{n}-\frac{p^{2}}{M},\boldsymbol{q}\right)\langle n|\hat{\boldsymbol{s}}_i|\boldsymbol{p}\rangle\langle \boldsymbol{p} |{\hat{\boldsymbol{s}}_j}|n\rangle\\
& \times \left(f_n\left(t, \boldsymbol{R},\boldsymbol{P}\right)-e^{-\beta\left(E_{n}-\frac{p^{2}}{M}\right) }f_o\left(t, \boldsymbol{R},\boldsymbol{P},\boldsymbol{0},\boldsymbol{p}\right) \right)
\end{split}  \label{boltz-singlet-no-recoil-small-dipol-2},
\end{equation}

\begin{equation}
\begin{split}
\left[\frac{\partial}{\partial t}+\frac{\boldsymbol{P}}{2M}.\nabla_{\boldsymbol{R}}+2\frac{\boldsymbol{k}}{M}.\nabla_{\boldsymbol{r}}\right]&f_o\left(t, \boldsymbol{R},\boldsymbol{P},\boldsymbol{r},\boldsymbol{k}\right) \\&=\frac{1}{2N_C}\delta_{ij}\delta\left(\boldsymbol{r}\right)\sum_{n}\int_{\boldsymbol{q}}\Delta_{\rm pNRQCD}\left(E_{n}-\frac{k^{2}}{M},\boldsymbol{q}\right)\langle n|\hat{\boldsymbol{s}}_i|\boldsymbol{k}\rangle\langle\boldsymbol{k}|\hat{\boldsymbol{s}}_j|n\rangle
\\& \times \left(f_n\left(t, \boldsymbol{R},\boldsymbol{P}\right)-e^{-\beta\left(E_{n}-\frac{k^{2}}{M}\right) }f_o\left(t, \boldsymbol{R},\boldsymbol{P},\boldsymbol{0},\boldsymbol{k}\right)\right)
\\&+\frac{N_c^2-4}{4 N_c}\delta_{ij}\delta\left(\boldsymbol{r}\right)\int_{\boldsymbol{q}\boldsymbol{p}}\Delta_{\rm pNRQCD}\left(\frac{p^{2}}{M}-\frac{k^{2}}{M},\boldsymbol{q}\right)\langle \boldsymbol{k}|\hat{\boldsymbol{s}}_i|\boldsymbol{p}\rangle\langle \boldsymbol{p}|\hat{\boldsymbol{s}}_j|\boldsymbol{k}\rangle\\
& \times \left(1-e^{-\beta\left(\frac{p^{2}}{M}-\frac{k^{2}}{M}\right) }\right)
f_o\left(t, \boldsymbol{R},\boldsymbol{P},\boldsymbol{0},\boldsymbol{k}\right)
\end{split}.\label{boltz-octet-no-recoil-small-dipole-2}
\end{equation}

As established in Sec.~\ref{section-NRQCD-Vs-pNRQCD}, the pNRQCD ULE coincides with the small-dipole limit of the NRQCD ULE. Consequently, the transport equations obtained here can equivalently be viewed as the semiclassical limit of the pNRQCD ULE in the small-dipole regime.

\subsection{Comparison with previous works}

We now compare Eqs.~(\ref{boltz-singlet-no-recoil-small-dipol-2})--(\ref{boltz-octet-no-recoil-small-dipole-2}) with the Boltzmann equations derived by Yao \textit{et al.} in Ref.~\cite{Yao:2018nmy}; see also Refs.~\cite{Yao:2021lus,yao2021semiclassical-transport,akamatsu2022quarkonium}. The latter were derived within the pNRQCD framework from the Davies secular equation and therefore rely on the rotating-wave approximation (RWA). The applicability of the RWA is limited when the relevant energy gaps become small. This issue is particularly important for in-medium quarkonium, since the octet scattering states form a continuum and therefore allow transitions with arbitrarily small energy differences~\cite{daddi2024,akamatsu2022dynamics}. In contrast, the transport equations derived here follow directly from the universal Lindblad equation without invoking the RWA, providing a natural framework for assessing the impact of this approximation on the semiclassical description.

The comparison performed below is independent of the explicit form of the medium correlator. Since the latter depends on the model used to describe the QGP rather than on the underlying quantum framework, we keep it in its general form, denoted by $\Delta_{\mathrm{pNRQCD}}$, and compare only the structural form of the transport equations.\footnote{In Ref.~\cite{Yao:2018nmy}, the corresponding medium correlator is evaluated explicitly for a weakly coupled QGP using a leading-order thermal gluon propagator. We leave it in its general form here in order to focus on the structure of the transport equations.} This allows us to separate differences arising from the derivation of the transport equations from those associated with a particular choice of medium correlator.

A comparison between the small-dipole limit of our Boltzmann equations, Eqs.~(\ref{boltz-singlet-no-recoil-small-dipol-2})--(\ref{boltz-octet-no-recoil-small-dipole-2}), and the corresponding equations of Ref.~\cite{Yao:2018nmy} shows that the two sets of equations are nearly identical, differing only in a few minor respects.

We first compare the singlet Boltzmann equation~(\ref{boltz-singlet-no-recoil-small-dipol-2}) with its counterpart derived by Yao \textit{et al.}~\cite{Yao:2018nmy,Yao:2021lus}, which can be written as follows~\cite{Yao:2021lus,akamatsu2022quarkonium}:\footnote{Since recoil effects were neglected in the derivation of Eqs.~(\ref{boltz-singlet-no-recoil-small-dipol-2})--(\ref{boltz-octet-no-recoil-small-dipole-2}), the corresponding approximation is also adopted when presenting the equations of Ref.~\cite{Yao:2018nmy}.}

\begin{equation}
\begin{split}
\left[\frac{\partial}{\partial t}+\frac{\boldsymbol{P}}{2M}.\nabla_{\boldsymbol{R}}\right]f_n\left(t, \boldsymbol{R},\boldsymbol{P}\right)=&-{C_{F}}\delta_{ij}\int_{\boldsymbol{q}\boldsymbol{p}}\Delta_{\rm pNRQCD}\left(\frac{p^{2}}{M}-E_{n},\boldsymbol{q}\right)\langle n|\hat{\boldsymbol{s}}_i|\boldsymbol{p}\rangle\langle \boldsymbol{p} |{\hat{\boldsymbol{s}}_j}|n\rangle\\
& \times \left(f_n\left(t, \boldsymbol{R},\boldsymbol{P}\right)-e^{-\beta\left(\frac{p^{2}}{M}-E_{n}\right) }f_o\left(t, \boldsymbol{R},\boldsymbol{P},\boldsymbol{0},\boldsymbol{p}\right) \right)
\end{split}  \label{boltz-singlet-yao},
\end{equation}

Several differences between the two formulations should be noted. First, the fields present in the Hamiltonian (3.1) were redefined in Ref.~\cite{Yao:2018nmy} to ensure gauge invariance and to account for the case of a strongly coupled QGP. For simplicity, we do not employ these redefinitions here. Nevertheless, they do not modify the structure of the transport equations relevant for the present comparison.

Second, the energy arguments of the QGP correlator and the Boltzmann factor, namely $\Delta(E)$ and $e^{-\beta E}$, differ only by an overall minus sign. This sign is purely conventional and originates from the use of opposite Fourier-transform conventions, namely, $e^{-iq_0t}$ versus $e^{iq_0t}$. Consequently, the energy-conservation delta functions take the forms $\delta(\frac{p^{2}}{M}-E_{n}+q_0)$ and $\delta(\frac{p^{2}}{M}-E_{n}-q_0)$, respectively. 

The same reasoning applies to the octet Boltzmann equations. In this case, however, our result contains an additional term compared with the corresponding equation of Ref.~\cite{Yao:2018nmy}; see Eq.~(4.58b) of Ref.~\cite{akamatsu2022quarkonium}. This term corresponds to the second collision term in Eq.~(\ref{boltz-octet-no-recoil-small-dipole-2}), namely
\begin{equation}
\begin{split}
&\frac{N_c^2-4}{4 N_c}\delta_{ij}\delta\left(\boldsymbol{r}\right)\int_{\boldsymbol{q}\boldsymbol{p}}\Delta_{\rm pNRQCD}\left(\frac{p^{2}}{M}-\frac{k^{2}}{M},\boldsymbol{q}\right)\langle \boldsymbol{k}|\hat{\boldsymbol{s}}_i|\boldsymbol{p}\rangle\langle \boldsymbol{p}|\hat{\boldsymbol{s}}_j|\boldsymbol{k}\rangle\\
& \times \left(1-e^{-\beta\left(\frac{p^{2}}{M}-\frac{k^{2}}{M}\right) }\right)
f_o\left(t, \boldsymbol{R},\boldsymbol{P},\boldsymbol{0},\boldsymbol{k}\right)
\end{split},
\end{equation}
which describes transitions within the continuum of octet scattering states, as indicated by the dipole transition matrix elements
$\langle\boldsymbol{k}|\hat{\boldsymbol{s}}_i|\boldsymbol{p}\rangle
\langle\boldsymbol{p}|\hat{\boldsymbol{s}}_j|\boldsymbol{k}\rangle$.
This differs from the first collision term, which describes transitions between the singlet and octet sectors.

The origin of this difference can be traced to the secular approximation underlying the Davies master equation. More generally, the system--medium interaction Hamiltonian can be written schematically as
\begin{equation}
H_{\mathrm{int}}=\sum_i A_i\otimes B_i,
\end{equation}
where $A_i$ and $B_i$ act on the system and medium Hilbert spaces, respectively. In the interaction picture, the system operators can be decomposed into components associated with the Bohr frequencies,
\begin{equation}
A_i(t)=\sum_{\omega}e^{-i\omega t}A_i(\omega),
\end{equation}
where $A_i(\omega)$ connects system states whose energy difference is $\omega$. The second-order master equation consequently contains terms proportional to
\begin{equation}
e^{i(\omega'-\omega)t}
A_i(\omega)\rho A_j^\dagger(\omega').
\end{equation}
The rotating-wave, or secular, approximation neglects the terms with $\omega'\neq\omega$, assuming that the corresponding phase factors oscillate rapidly and average to zero on the relaxation time scale. This requires the relevant transition energies to be sufficiently well separated, or equivalently that the differences $|\omega'-\omega|$ be large compared with the inverse relaxation time~\cite{akamatsu2022quarkonium,mozgunov2020completely,davidovic2020completely}.

This condition is not uniformly satisfied in the octet continuum. Indeed, the octet scattering states have continuous energies,
\begin{equation}
E_{\boldsymbol{k}}=\frac{\boldsymbol{k}^{2}}{M},
\end{equation}
so that the corresponding transition energies form a continuous set. Consequently, distinct transitions can become arbitrarily close in energy,
\begin{equation}
|\omega'-\omega|\longrightarrow0,
\end{equation}
and their relative phase is no longer rapidly oscillating on the relaxation time scale. The secular approximation is therefore not uniformly justified in the octet sector. Contributions from nearly degenerate transitions, which are discarded in the Davies construction, can survive in the universal Lindblad equation. In the semiclassical limit, these contributions give rise to the additional octet--octet collision term in Eq.~(\ref{boltz-octet-no-recoil-small-dipole-2}), thereby explaining its absence from the Boltzmann equation derived by Yao \textit{et al.}~\cite{Yao:2018nmy}. Thus, the additional term is not introduced as an independent ingredient of the transport description, but follows directly from retaining the nearly degenerate transitions in the universal Lindblad framework.

This additional collision term may influence the dynamics of in-medium quarkonia, in particular the approach to equilibrium and the associated thermalization time scale. Quantifying its phenomenological impact requires solving the coupled singlet--octet transport equations and is left for future work.
\section{Quantum corrections}
\label{section-quantum-corrections}

In the derivation of the singlet--octet Boltzmann equations presented in the previous section, only the leading-order term of the semiclassical expansion, Eq.~(\ref{gradient-expansion-main}), was retained. In this section, we evaluate the contribution from the next-to-leading-order term of this expansion. This provides the leading quantum correction to the semiclassical transport equations~\cite{haug2008quantum,yao2021semiclassical-transport}.

It is important to note that these quantum corrections arise only in the collision terms involving the octet Wigner function, namely the octet-to-singlet and octet-to-octet transition terms. This follows from the fact that the gradient expansion, Eq.~(\ref{gradient-expansion-main}), is performed only for the octet Wigner function, while no analogous expansion can be introduced for the singlet Wigner function.\footnote{The gradient expansion~(\ref{gradient-expansion-main}) relies on the dependence of the octet Wigner function on the continuous relative coordinate $\boldsymbol{r}$, about which the expansion is performed. By contrast, the singlet Wigner function is defined only for discrete bound states, $f_n(t,\boldsymbol{R},\boldsymbol{P})$, and therefore does not admit an analogous semiclassical expansion.}

We first derive in detail the leading quantum correction to the octet-to-singlet transition term given by Eq.~(\ref{194}). The corresponding correction to the octet-to-octet transition term can be obtained following the same procedure. Throughout this derivation, we closely follow the approach of Ref.~\cite{yao2021semiclassical-transport}.

Retaining the next-to-leading-order term in the semiclassical expansion, Eq.~(\ref{gradient-expansion-main}), and subsequently performing the Wigner transformation yields
\begin{equation}
    \begin{split}&C_{F}\int_{\boldsymbol{q}\boldsymbol{p}\boldsymbol{p}^{\prime}\boldsymbol{r}}e^{-i\left(\boldsymbol{p}^\prime-\boldsymbol{p}\right).\boldsymbol{r}} g\left(\frac{p^{2}}{M}-E_{n},\boldsymbol{q}\right)g\left(\frac{p^{\prime2}}{M}-E_{n},\boldsymbol{q}\right)\\
    & \times\langle n|S_{\boldsymbol{q}.\hat{\boldsymbol{s}}}|\boldsymbol{p}^{\prime}\rangle\langle \boldsymbol{p}|S_{\boldsymbol{q}.\hat{\boldsymbol{s}}}|n\rangle  \left(\boldsymbol{r}-\boldsymbol{r}_0\right).\nabla_{\boldsymbol{r}_0}f_o\left(t, \boldsymbol{R},\boldsymbol{P},\boldsymbol{r}_0,\tilde{\boldsymbol{p}}\right)
    \end{split},
\end{equation}
we can set, for simplicity, $\boldsymbol{r_0}=0$ and rewrite this expression as follows
\begin{equation}
    \begin{split}&-iC_{F}\int_{\boldsymbol{q}\boldsymbol{p}\boldsymbol{p}^{\prime}}g\left(\frac{p^{2}}{M}-E_{n},\boldsymbol{q}\right)g\left(\frac{p^{\prime2}}{M}-E_{n},\boldsymbol{q}\right)\langle n|S_{\boldsymbol{q}.\hat{\boldsymbol{s}}}|\boldsymbol{p}^{\prime}\rangle\langle \boldsymbol{p}|S_{\boldsymbol{q}.\hat{\boldsymbol{s}}}|n\rangle\\
    & \times \left[ \frac{\nabla_{\boldsymbol{p}}-\nabla_{\boldsymbol{p}^\prime}}{2}e^{-i\left(\boldsymbol{p}^\prime-\boldsymbol{p}\right).\boldsymbol{r}} \right].\nabla_{\boldsymbol{r}_0}f_o\left(t, \boldsymbol{R},\boldsymbol{P},\boldsymbol{r}_0,\tilde{\boldsymbol{p}}\right)|_{\boldsymbol{r}_0=0}\\
    =&-iC_{F}\int_{\boldsymbol{q}\boldsymbol{p}\boldsymbol{p}^{\prime}}g\left(\frac{p^{2}}{M}-E_{n},\boldsymbol{q}\right)g\left(\frac{p^{\prime2}}{M}-E_{n},\boldsymbol{q}\right)\langle n|S_{\boldsymbol{q}.\hat{\boldsymbol{s}}}|\boldsymbol{p}^{\prime}\rangle\langle \boldsymbol{p}|S_{\boldsymbol{q}.\hat{\boldsymbol{s}}}|n\rangle\\
    & \times \left[ \frac{\nabla_{\boldsymbol{p}}-\nabla_{\boldsymbol{p}^\prime}}{2}\delta\left(\boldsymbol{p}^\prime-\boldsymbol{p}\right) \right].\nabla_{\boldsymbol{r}_0}f_o\left(t, \boldsymbol{R},\boldsymbol{P},\boldsymbol{r}_0,\tilde{\boldsymbol{p}}\right)|_{\boldsymbol{r}_0=0}
    \end{split},
\end{equation}
where in the second equality we have performed the integration over $\boldsymbol{r}$.

We next integrate by parts over the momenta $\boldsymbol{p}$ and $\boldsymbol{p}^\prime$:
\begin{equation}
    \begin{split}&iC_{F}\int_{\boldsymbol{q}\boldsymbol{p}\boldsymbol{p}^{\prime}}\delta\left(\boldsymbol{p}^\prime-\boldsymbol{p}\right)\left[ \frac{\nabla_{\boldsymbol{p}}-\nabla_{\boldsymbol{p}^\prime}}{2} \right]g\left(\frac{p^{2}}{M}-E_{n},\boldsymbol{q}\right)g\left(\frac{p^{\prime2}}{M}-E_{n},\boldsymbol{q}\right)\\
    & \times \langle n|S_{\boldsymbol{q}.\hat{\boldsymbol{s}}}|\boldsymbol{p}^{\prime}\rangle\langle \boldsymbol{p}|S_{\boldsymbol{q}.\hat{\boldsymbol{s}}}|n\rangle \nabla_{\boldsymbol{r}_0}f_o\left(t, \boldsymbol{R},\boldsymbol{P},\boldsymbol{r}_0,\tilde{\boldsymbol{p}}\right)|_{\boldsymbol{r}_0=0}
    \end{split},
\end{equation}

After performing the integration enforced by the Dirac delta function, several terms cancel, leaving the following expression for the quantum correction to the octet-to-singlet transition term:
\begin{equation}
-i\frac{C_F}{2}\int_{\boldsymbol{q}\boldsymbol{p}}\left[g\left(\frac{p^{2}}{M}-E_{n},\boldsymbol{q}\right)\right]^2|\langle n|S_{\boldsymbol{q}.\hat{\boldsymbol{s}}}|\boldsymbol{p}\rangle|^2 \left(\frac{\nabla_{\boldsymbol{p}} \langle n|S_{\boldsymbol{q}.\hat{\boldsymbol{s}}}|\boldsymbol{p}\rangle}{ \langle n|S_{\boldsymbol{q}.\hat{\boldsymbol{s}}}|\boldsymbol{p}\rangle}- \frac{\nabla_{\boldsymbol{p}}\langle \boldsymbol{p}|S_{\boldsymbol{q}.\hat{\boldsymbol{s}}}|n\rangle}{\langle \boldsymbol{p}|S_{\boldsymbol{q}.\hat{\boldsymbol{s}}}|n\rangle} \right) .\nabla_{\boldsymbol{r}_0}f_o\left(t, \boldsymbol{R},\boldsymbol{P},\boldsymbol{r}_0,{\boldsymbol{p}}\right)|_{\boldsymbol{r}_0=0}.
\end{equation}

In the small-dipole limit, the quantum correction obtained above can be compared with the corresponding result of Yao \textit{et al.}~\cite{yao2021semiclassical-transport}. The two expressions are found to be in close agreement, apart from the differences already identified in the leading-order transport equations. This agreement provides an additional consistency check of the semiclassical expansion and of the relation between the two approaches.

Including the leading quantum correction, the singlet Boltzmann equation~(\ref{boltz-singlet-no-recoil}) becomes
\begin{equation}
  \begin{split}
    \left[\frac{\partial}{\partial t}+\frac{\boldsymbol{P}}{2M}.\nabla_{\boldsymbol{R}}\right]f_n\left(t, \boldsymbol{R},\boldsymbol{P}\right)=&-{C_{F}}\int_{\boldsymbol{q}\boldsymbol{p}}\Delta^>\left(E_{n}-\frac{p^{2}}{M},\boldsymbol{q}\right)|\langle n|S_{\boldsymbol{q}.\hat{\boldsymbol{s}}}|\boldsymbol{p}\rangle|^2\\
        & \times \left[f_n\left(t, \boldsymbol{R},\boldsymbol{P}\right)-e^{-\beta\left(E_{n}-\frac{p^{2}}{M}\right) }f_o\left(t, \boldsymbol{R},\boldsymbol{P},\boldsymbol{0},\boldsymbol{p}\right)\right)\\
      & -i\frac{C_{F}}{2}\int_{\boldsymbol{q}\boldsymbol{p}}\Delta^>\left(\frac{p^{2}}{M}-E_{n},\boldsymbol{q}\right)|\langle n|S_{\boldsymbol{q}.\hat{\boldsymbol{s}}}|\boldsymbol{p}\rangle|^2\\
        &\times \left(\frac{\nabla_{\boldsymbol{p}} \langle n|S_{\boldsymbol{q}.\hat{\boldsymbol{s}}}|\boldsymbol{p}\rangle}{ \langle n|S_{\boldsymbol{q}.\hat{\boldsymbol{s}}}|\boldsymbol{p}\rangle}- \frac{\nabla_{\boldsymbol{p}}\langle \boldsymbol{p}|S_{\boldsymbol{q}.\hat{\boldsymbol{s}}}|n\rangle}{\langle \boldsymbol{p}|S_{\boldsymbol{q}.\hat{\boldsymbol{s}}}|n\rangle} \right) .\nabla_{\boldsymbol{r}_0}f_o\left(t, \boldsymbol{R},\boldsymbol{P},\boldsymbol{r}_0,{\boldsymbol{p}}\right)|_{\boldsymbol{r}_0=0}
  \end{split}  \label{boltz-singlet-qm},
\end{equation}
where, in the quantum correction, the jump correlator function has been expressed in terms of the QGP thermal correlator using Eq.~(\ref{eq:10}).

The same procedure can be applied to the octet Boltzmann equation~(\ref{boltz-octet-no-recoil}) to derive and incorporate its quantum corrections. The resulting expression is lengthy and is therefore not displayed here. 

\section{Conclusions}
\label{Conclusions}

In this work, we extended the results of Ref.~\cite{daddigeneral}, where a set of coupled singlet--octet universal Lindblad equations (ULEs) was derived within NRQCD for the description of in-QGP quarkonia. We derived the corresponding pNRQCD ULEs and showed that they coincide with the small-dipole limit of the NRQCD ULEs, thereby establishing the small-dipole limit as a common framework for comparing the equations derived within the two effective field theories.

We then explored the semiclassical limit of the singlet--octet ULEs derived in Ref.~\cite{daddigeneral} and obtained the corresponding coupled singlet--octet Boltzmann equations, thereby establishing a direct connection between the universal Lindblad framework and semiclassical transport. To our knowledge, this constitutes the first derivation of Boltzmann transport equations directly from the universal Lindblad framework. The resulting equations are valid beyond the small-dipole approximation and therefore provide a more general framework for the semiclassical description of quarkonium transport in the QGP.

We then considered the small-dipole limit of the resulting transport equations and compared them with the Boltzmann equations derived by Yao \textit{et al.}~\cite{Yao:2018nmy}. The latter were obtained within pNRQCD from the Davies secular equation and therefore rely on the rotating-wave approximation (RWA), whereas the transport equations derived here follow from the universal Lindblad equation without invoking this approximation. The singlet equations are found to be in near-complete agreement. By contrast, the octet equation contains an additional collision term describing transitions among octet scattering states with arbitrarily small energy differences. This contribution originates from the continuous octet spectrum, for which the secular approximation is not uniformly justified, and is absent from the transport equations of Ref.~\cite{Yao:2018nmy}. The universal Lindblad framework retains contributions from nearly degenerate transitions that are discarded in the RWA-based treatment.

Finally, we derived the leading quantum correction to the singlet Boltzmann equation. In the small-dipole limit, this correction is found to be in close agreement with the corresponding result of Ref.~\cite{yao2021semiclassical-transport}, providing an additional consistency check of the semiclassical expansion.

The general Boltzmann equations derived in this work also have direct phenomenological relevance. The validity of the transport equations beyond the small-dipole regime is also relevant from a phenomenological perspective, particularly for excited quarkonium states and for stages of the evolution in which the heavy-quark pair reaches large separations. The present framework can therefore describe the evolution from compact bound configurations to widely separated and unbound $Q\bar Q$ states within the same transport description, allowing the dynamics through and beyond the dissociation region to be studied without introducing a separate small-dipole restriction. This provides a natural starting point for future phenomenological studies, in particular to quantify the impact of the additional octet collision term and of the quantum corrections on realistic simulations of quarkonium transport in heavy-ion collisions. It would also be desirable to compare the full quantum singlet--octet ULEs with their semiclassical counterparts, extending the quantum--semiclassical comparison of Ref.~\cite{Daddi-Hammou:2025hdz} beyond the quantum Brownian regime and thereby providing a broader assessment of the validity of the semiclassical approximation.

\acknowledgments
The authors acknowledge fruitful discussions with Thierry Gousset and Jean-Paul Blaizot. 
\appendix

\section{Projection onto color space of the pNRQCD ULE}\label{Appendinx-pNRQCD-Color-Projection}
In order to perform the projection in color space, we utilize the same color diagonal density matrix that was employed in Section \ref{NRQCD-ULE}, that is  given by \cite{blaizot2018quantum,blaizot2018approach}  
\begin{equation}
\rho =\rho_{s}\left|s\right\rangle \left\langle s\right|+\rho_{o}\underset{c}{\sum}\left|c\right\rangle \left\langle c\right|.
\end{equation}

So our current objective is to project the derived pNRQCD ULE (\ref{eq:22-1}) on the singlet-octet basis and  write the singlet-octet equations as follows 

\begin{equation}
\frac{d\rho_{s}\left(t\right)}{dt}=-i\left[H_{s}+\Lambda_{s} ,\rho_{s}\left(t\right)\right]+\mathcal{L}_{ss}\left(t\right)\rho_{s}\left(t\right)+\mathcal{L}_{so}\left(t\right)\rho_{o}\left(t\right)\label{eq:singlet equation22-2},
\end{equation}
\begin{equation}
\frac{d\rho_{o}\left(t\right)}{dt}=-i\left[H_{o}+\Lambda_{o},\rho_{o}\left(t\right)\right]+\mathcal{L}_{os}\left(t\right)\rho_{s}\left(t\right)+\mathcal{L}_{oo}\left(t\right)\rho_{o}\left(t\right)\label{octet eqaution22-2}.
\end{equation}

\subsection{The singlet  ULE}\label{pNQRCD Singlet master equation subsection}

The projection of the lefthand side of Eq. (\ref{eq:22-1})   yields 
\begin{equation}
\langle s|\frac{d\rho \left(t\right)}{dt}|s\rangle =\frac{d\rho_{s}\left(t\right)}{dt}.
\end{equation}

For the sake of clarity,  the projection of the  righthand side of Eq. (\ref{eq:22-1}) can be considered
term by term, as was done with the NRQCD case. 
 Thus, let us begin with the Liouville superoperator.

By projecting  the first term in the singlet-singlet matrix element $\mathcal{L}_{ss}\left(t\right)\rho_{s}\left(t\right)$, i.e.   $L_{i}^{\dagger}\left(\boldsymbol{y}\right)L_{i}\left(\boldsymbol{y}\right)\rho \left(t\right) $, on singlet states,  we obtain\footnote{To lighten up the notation,  we omit the integration over  $v$, and  $v^\prime$ in the following computations, but we  restore them in the final expressions. }

\begin{equation}
\begin{array}{ccl}
\langle s|L_{i}^{\dagger}\left(\boldsymbol{y}\right)L_{i}\left(\boldsymbol{y}\right)\rho \left(t\right)|s\rangle  & \equiv& g\left(v-t,\boldsymbol{R}-\boldsymbol{y}\right)g\left(t-v^{\prime},\boldsymbol{y}-\boldsymbol{R}^\prime\right)\\
&& \times \langle s|U\left(t-v\right)V_{i}^{a}U^{\dagger}\left(t-v\right)U\left(t-v^{\prime}\right)V_{i}^{a}U^{\dagger}\left(t-v^{\prime}\right)\rho \left(t\right)|s\rangle \\
\\
 & = & g\left(v-t,\boldsymbol{R}-\boldsymbol{y}\right)g\left(t-v^{\prime},\boldsymbol{y}-\boldsymbol{R}^\prime\right)\\
 && \times U_{s}\left(t-v\right)\langle s|V_{i}^{a}U^{\dagger}\left(t-v\right)U\left(t-v^{\prime}\right)V_{i}^{a}|s\rangle U_{s}^{\dagger}\left(t-v^{\prime}\right)\rho_{s}\left(t\right)\\

\end{array},
\end{equation}
then, using $V_{i}^{a}|s\rangle =\frac{r_{i}}{\sqrt{2N_{c}}}|a\rangle$ and $\left\langle a|a\right\rangle =N_{c}^{2}-1$, the last  equation yields 
\begin{equation}
\begin{array}{ccl}
\langle s|L_{i}^{\dagger}\left(\boldsymbol{y}\right)L_{i}\left(\boldsymbol{y}\right)\rho \left(t\right)|s\rangle 
 & = & \frac{N_{c}^{2}-1}{2N_{c}}g\left(v-t,\boldsymbol{R}-\boldsymbol{y}\right)g\left(t-v^{\prime},\boldsymbol{y}-\boldsymbol{R}^\prime\right)\\
&& \times U_{s}\left(t-v\right)r_{i}U_{o}^{\dagger}\left(t-v\right)U_{o}\left(t-v^{\prime}\right)r_{i}^{\prime}U_{s}^{\dagger}\left(t-v^{\prime}\right)\rho_{s}\left(t\right)\\
\end{array}.
\end{equation}

Following the same steps for the second term $\langle s|\rho \left(t\right)L_{i}^{\dagger}\left(\boldsymbol{y}\right)L_{i}\left(\boldsymbol{y}\right)|s\rangle $, 
we get the following final  expression for Singlet-Singlet matrix element
\begin{equation}
\begin{array}{ccl}
\\
\mathcal{L}_{ss}\left(t\right)\rho_{s}\left(t\right) & = & -\frac{C_{F}}{2}\int_{v,v^{\prime},\boldsymbol{y}}g\left(v-t,\boldsymbol{R}-\boldsymbol{y}\right)g\left(t-v^{\prime},\boldsymbol{y}-\boldsymbol{R}^\prime\right)\\
&& \times \left\{ U_{s}\left(t-v\right)r_{i}U_{o}^{\dagger}\left(t-v\right)U_{o}\left(t-v^{\prime}\right)r_{i}^{\prime}U_{s}^{\dagger}\left(t-v^{\prime}\right),\rho_{s}\left(t\right)\right\} 
\end{array}.\\
\end{equation}

By Fourier transforming this last equation, accroding to 
\begin{equation}
g\left(t-v,\boldsymbol{y}-\boldsymbol{R}\right)=\int_{\boldsymbol{q}}e^{i\boldsymbol{q}\left(\boldsymbol{y}-\boldsymbol{R}\right)}g\left(t-v,\boldsymbol{q}\right)\label{eq:91},
\end{equation}
 \begin{equation}
g\left(t-v,\boldsymbol{y}-\boldsymbol{R}\right)=\int_{q_{0}}e^{-iq_{0}\left(t-v\right)}g\left(q_{0},\boldsymbol{y}-\boldsymbol{R}\right)\label{eq:92-1}.
\end{equation}
we get\footnote{In this equation, we consider the center of mass to be integrated out. This is achieved by setting $\boldsymbol{R}=\boldsymbol{R}^{\prime}$, as done in the   case of NRQCD in \cite{daddigeneral}. This results in  
\begin{equation}
\int_{\boldsymbol{z}}g\left(v-t,\boldsymbol{R-z}\right)g\left(t-v^{\prime},\boldsymbol{y}-\boldsymbol{R}^\prime\right)\rightarrow\int_{\boldsymbol{q}}e^{i\boldsymbol{q}\left(\boldsymbol{R}-\boldsymbol{R}^{\prime}\right)}g\left(v-t,\boldsymbol{q}\right)g\left(t-v^{\prime},\boldsymbol{q}\right)\overset{\boldsymbol{R}=\boldsymbol{R}^{\prime}}{\Longrightarrow}\int_{\boldsymbol{q}}g\left(v-t,\boldsymbol{q}\right)g\left(t-v^{\prime},\boldsymbol{q}\right)
\end{equation}
}
\begin{equation}
\begin{array}{ccl}\\
\tilde{\mathcal{L}}_{ss}\left(t\right)\tilde{\rho}_{s}\left(t\right) & = & -\frac{C_{F}}{2}\int_{vv^{\prime}q_{0}q_{0}^{\prime}\boldsymbol{q}}e^{i\left(q_{0}-q_{0}^{\prime}\right)t}e^{-iq_{0}v+iq_{0}^{\prime}v^{\prime}}g\left(q_{0},\boldsymbol{q}\right)g\left(q_{0}^{\prime},\boldsymbol{q}\right)\\
&& \times \left\{ \tilde{U}_{s}\left(t-v\right)r_{i}\tilde{U}_{o}^{\dagger}\left(t-v\right)\tilde{U}_{o}\left(t-v^{\prime}\right)r_{i}^{\prime}\tilde{U}_{s}^{\dagger}\left(t-v^{\prime}\right),\tilde{\rho}_{s}\left(t\right)\right\} 
\end{array}\label{eq:187-2}\\.
\end{equation}

The octet to singlet transition element   can be obtained by projecting the term  $L\left(\boldsymbol{y}\right)\rho \left(t\right)L^{\dagger}\left(\boldsymbol{y}\right)$ on singlet states as follows: 
\begin{equation}
\begin{array}{ccl}
\langle s|L\left(\boldsymbol{y}\right)\rho \left(t\right)L^{\dagger}\left(\boldsymbol{y}\right)|s\rangle  & \equiv& g\left(v-t,\boldsymbol{R}-\boldsymbol{y}\right)g\left(t-v^{\prime},\boldsymbol{y}-\boldsymbol{R}^\prime\right)\\
&& \times \langle s|U\left(t-v\right)V_{i}^{a}U^{\dagger}\left(t-v\right)\rho \left(t\right)U\left(t-v^{\prime}\right)V_{i}^{a}U^{\dagger}\left(t-v^{\prime}\right)|s\rangle \\
\\
 & =&\frac{N_{c}^{2}-1}{2N_{c}}g\left(v-t,\boldsymbol{R}-\boldsymbol{y}\right)g\left(t-v^{\prime},\boldsymbol{y}-\boldsymbol{R}^\prime\right)\\
 && \times U_{s}\left(t-v\right)r_{i}U_{o}^{\dagger}\left(t-v\right)\rho_{o}\left(t\right)U_{o}\left(t-v^{\prime}\right)r^\prime_{i}U_{s}^{\dagger}\left(t-v^{\prime}\right)
\end{array}.\label{8.202-2}
\end{equation}

The final expression of $\tilde{\mathcal{L}}_{so}\left(t\right)\rho_{o}\left(t\right)$  is obtained by applying the Fourier transform to the jump correlator functions, using Eq. (\ref{eq:91}) and Eq.(\ref{eq:92-1}).
This results in the following expression:
\begin{equation}
\begin{array}{ccl}
\tilde{\mathcal{L}}_{so}\left(t\right)\tilde{\rho}_{o}\left(t\right) & = & C_{F}\int_{vv^{\prime}q_{0}q_{0}^{\prime}\boldsymbol{q}}e^{i\left(q_{0}-q_{0}^{\prime}\right)t}e^{-iq_{0}v+iq_{0}^{\prime}v^{\prime}}g\left(q_{0},\boldsymbol{q}\right)g\left(q_{0}^{\prime},\boldsymbol{q}\right)\\
&& \times \tilde{U}_{s}\left(t-v\right)r_{i}\tilde{U}_{o}^{\dagger}\left(t-v\right)\tilde{\rho}_{o}\left(t\right)\tilde{U}_{o}\left(t-v^{\prime}\right)r_{i}^{\prime}\tilde{U}_{s}^{\dagger}\left(t-v^{\prime}\right)
\end{array}
\end{equation}

The singlet  Lamb-shift  term is given by
\footnote{ This expression  can be easily obtained by following the steps that  led to Eq. (\ref{eq:187-2}).} 
\begin{equation}
\begin{array}{ccl}
\tilde{\Lambda}_{s}  
 & = & \frac{C_{F}}{2i}\int_{vv^{\prime}q_{0}q_{0}^{\prime}\boldsymbol{q}}e^{i\left(q_{0}-q_{0}^{\prime}\right)t}e^{-iq_{0}v+iq_{0}^{\prime}v^{\prime}}g\left(q_{0},\boldsymbol{q}\right)g\left(q_{0}^{\prime},\boldsymbol{q}\right)\\
 && \times \tilde{U}_{s}\left(t-v\right)r_{i}\tilde{U}_{o}^{\dagger}\left(t-v\right)\tilde{U}_{o}\left(t-v^{\prime}\right)r_{i}^{\prime}\tilde{U}_{s}^{\dagger}\left(t-v^{\prime}\right)
\end{array}.
\end{equation}

\subsection{The octet ULE}


Projecting the pNRQCD ULE into the octet states, we get for the lefthand
side of Eq.  (\ref{eq:22-1}) 
\begin{equation}
\left\langle a\left|\frac{d\rho \left(t\right)}{dt}\right|a\right\rangle =\left(N_{c}^{2}-1\right)\frac{d\rho_{o}\left(t\right)}{dt}\longrightarrow\frac{d\rho_{o}\left(t\right)}{dt}.
\end{equation}

Since the factor  $N_{c}^{2}-1$ shows up 
 on both sides of Eq. (\ref{eq:22-1}), after projection, this factor will dissapear in the final expression of the octet pNRQCD ULE.

The projection of the righthand side of  Eq. (\ref{eq:22-1}) is analogous to the singlet case that was previously presented, thus, we report  directly the final expressions.

\begin{equation}
\protect\begin{array}{ccl}
\protect\\
\tilde{\mathcal{L}}_{os}\left(t\right)\tilde{\rho}_{s}\left(t\right) & = & \frac{1}{2N_{c}}\int_{vv^{\prime}q_{0}q_{0}^{\prime}\boldsymbol{q}}e^{i\left(q_{0}-q_{0}^{\prime}\right)t}e^{-iq_{0}v+iq_{0}^{\prime}v^{\prime}}g\left(q_{0},\boldsymbol{q}\right)g\left(q_{0}^{\prime},\boldsymbol{q}\right)\\
&& \times \tilde{U}_{o}\left(t-v\right)r_{i}\tilde{U}_{s}^{\dagger}\left(t-v\right)\tilde{\rho}_{s}\left(t\right)\tilde{U}_{s}\left(t-v^{\prime}\right)r_{i}^{\prime}\tilde{U}_{o}^{\dagger}\left(t-v^{\prime}\right)\protect
\end{array}\label{eq:206-2},
\end{equation}

\begin{equation}
\protect\begin{array}{ccl}
\protect
\tilde{\mathcal{L}}_{oo}\left(t\right)\tilde{\rho}_{o}\left(t\right) & = & \frac{N_{c}^{2}-4}{4N_{c}}\int_{vv^{\prime}q_{0}q_{0}^{\prime}\boldsymbol{q}}e^{i\left(q_{0}-q_{0}^{\prime}\right)t}e^{-iq_{0}v+iq_{0}^{\prime}v^{\prime}}g\left(q_{0},\boldsymbol{q}\right)g\left(q_{0}^{\prime},\boldsymbol{q}\right)\\
&& \times \tilde{U}_{o}\left(t-v\right)r_{i}\tilde{U}_{o}^{\dagger}\left(t-v\right)\tilde{\rho}_{o}\left(t\right)\tilde{U}_{o}\left(t-v^{\prime}\right)r_{i}^{\prime}\tilde{U}_{o}^{\dagger}\left(t-v^{\prime}\right)\protect\\
\\
 &  & -\frac{1}{4N_{c}}\int_{vv^{\prime}q_{0}q_{0}^{\prime}\boldsymbol{q}}e^{i\left(q_{0}-q_{0}^{\prime}\right)t}e^{-iq_{0}v+iq_{0}^{\prime}v^{\prime}}g\left(q_{0},\boldsymbol{q}\right)g\left(q_{0}^{\prime},\boldsymbol{q}\right)\\
 && \times \left\{ \tilde{U}_{o}\left(t-v\right)r_{i}\tilde{U}_{s}^{\dagger}\left(t-v\right)\tilde{U}_{s}\left(t-v^{\prime}\right)r_{i}^{\prime}\tilde{U}_{o}^{\dagger}\left(t-v^{\prime}\right),\tilde{\rho}_{o}\left(t\right)\right\} \\
 \\
 &  & -\frac{N_{c}^{2}-4}{8N_{c}}\int_{vv^{\prime}q_{0}q_{0}^{\prime}\boldsymbol{q}}e^{i\left(q_{0}-q_{0}^{\prime}\right)t}e^{-iq_{0}v+iq_{0}^{\prime}v^{\prime}}g\left(q_{0},\boldsymbol{q}\right)g\left(q_{0}^{\prime},\boldsymbol{q}\right)\\
&& \times\left\{ \tilde{U}_{o}\left(t-v\right)r_{i}\tilde{U}_{o}^{\dagger}\left(t-v\right)\tilde{U}_{o}\left(t-v^{\prime}\right)r_{i}^{\prime}\tilde{U}_{o}^{\dagger}\left(t-v^{\prime}\right),\tilde{\rho}_{o}\left(t\right)\right\} 
\end{array}\label{pNRQCD-octet-octet-element-2},
\end{equation}

\begin{equation}
\protect\begin{array}{ccl}
\protect
\tilde{\Lambda}_{o} & = & \frac{1}{4iN_{c}}\int_{vv^{\prime}q_{0}q_{0}^{\prime}\boldsymbol{q}}\text{sign}\left(v-v^{\prime}\right)e^{i\left(q_{0}-q_{0}^{\prime}\right)t}e^{-iq_{0}v+iq_{0}^{\prime}v^{\prime}}g\left(q_{0},\boldsymbol{q}\right)g\left(q_{0}^{\prime},\boldsymbol{q}\right)\\
&& \times \tilde{U}_{o}\left(t-v\right)r_{i}\tilde{U}_{s}^{\dagger}\left(t-v\right)\tilde{U}_{s}\left(t-v^{\prime}\right)r_{i}^{\prime}\tilde{U}_{o}^{\dagger}\left(t-v^{\prime}\right)\\
\\
 &  & +\frac{N_{c}^{2}-4}{8iN_{c}}\int_{vv^{\prime}q_{0}q_{0}^{\prime}\boldsymbol{q}}\text{sign}\left(v-v^{\prime}\right)e^{i\left(q_{0}-q_{0}^{\prime}\right)t}e^{-iq_{0}v+iq_{0}^{\prime}v^{\prime}}g\left(q_{0},\boldsymbol{q}\right)g\left(q_{0}^{\prime},\boldsymbol{q}\right)\\
 && \times \tilde{U}_{o}\left(t-v\right)r_{i}\tilde{U}_{o}^{\dagger}\left(t-v\right)\tilde{U}_{o}\left(t-v^{\prime}\right)r_{i}^{\prime}\tilde{U}_{o}^{\dagger}\left(t-v^{\prime}\right)
\end{array}\label{pNRQCD-octet-LS-2}.
\end{equation}

\section{The octet ULE expression}\label{Appendix-The octet ULE matrix elements}

The expressions of the various terms in the octet ULE (\ref{octet-eq-projected-on-energy}) are given by 
\begin{equation}
\begin{array}{ccl}
\langle \boldsymbol{k}|\left[\tilde{\Lambda}_{o} ,\tilde{\rho}_{o}\left(t\right)\right]|\boldsymbol{k}^{\prime}\rangle  & = & -\frac{1}{4}\frac{N_{c}^{2}-4}{N_{c}}\int_{\boldsymbol{q}\boldsymbol{p}\boldsymbol{p}^{\prime}}\text{P}\int_{q_{0}^{\prime}}\frac{g\left(q_{0}^{\prime}-\frac{p^{\prime2}}{M}+\frac{k^{2}}{M},\boldsymbol{q}\right)g\left(q_{0}^{\prime},\boldsymbol{q}\right)}{q_{0}^{\prime}-\frac{p^{\prime2}}{M}+\frac{p^{2}}{M}}\\
 &  & \times\langle \boldsymbol{k}|S_{\boldsymbol{q}.\hat{\boldsymbol{s}}}|\boldsymbol{p}\rangle \langle \boldsymbol{p}|S_{\boldsymbol{q}.\hat{\boldsymbol{s}}^{\prime}}|\boldsymbol{p}^{\prime}\rangle\langle \boldsymbol{p}^{\prime}|\tilde{\rho}_{o}\left(t\right)|\boldsymbol{k}^{\prime}\rangle \\
\\
 &  & +\frac{1}{4}\frac{N_{c}^{2}-4}{N_{c}}\int_{\boldsymbol{q}\boldsymbol{p}\boldsymbol{p}^{\prime}}\text{P}\int_{q_{0}^{\prime}}\frac{g\left(q_{0}^{\prime}-\frac{k^{\prime2}}{M}+\frac{p^{\prime2}}{M},\boldsymbol{q}\right)g\left(q_{0}^{\prime},\boldsymbol{q}\right)}{q_{0}^{\prime}-\frac{k^{\prime2}}{M}+\frac{p^{2}}{M}}\\
 &  & \times\langle \boldsymbol{k}|\tilde{\rho}_{o}\left(t\right)|\boldsymbol{p}^{\prime}\rangle\langle \boldsymbol{p}^{\prime}|S_{\boldsymbol{q}.\hat{\boldsymbol{s}}}|\boldsymbol{p}\rangle \langle \boldsymbol{p}|S_{\boldsymbol{q}.\hat{\boldsymbol{s}}^{\prime}}|\boldsymbol{k}^{\prime}\rangle  \\
\\
 &  &-\frac{N_{c}}{4}\int_{\boldsymbol{q}\boldsymbol{p}\boldsymbol{p}^{\prime}}\text{P}\int_{q_{0}^{\prime}}\frac{g\left(q_{0}^{\prime}-\frac{p^{\prime2}}{M}+\frac{k^{2}}{M},\boldsymbol{q}\right)g\left(q_{0}^{\prime},\boldsymbol{q}\right)}{q_{0}^{\prime}-\frac{p^{\prime2}}{M}+\frac{p^{2}}{M}}\\
 &  & \times\langle \boldsymbol{k}|C_{\boldsymbol{q}.\hat{\boldsymbol{s}}}|\boldsymbol{p}\rangle \langle \boldsymbol{p}|C_{\boldsymbol{q}.\hat{\boldsymbol{s}}^{\prime}}|\boldsymbol{p}^{\prime}\rangle\langle \boldsymbol{p}^{\prime}|\tilde{\rho}_{o}\left(t\right)|\boldsymbol{k}^{\prime}\rangle \\
\\
 &  & +\frac{N_{c}}{4}\int_{\boldsymbol{q}\boldsymbol{p}\boldsymbol{p}^{\prime}}\text{P}\int_{q_{0}^{\prime}}\frac{g\left(q_{0}^{\prime}-\frac{k^{\prime2}}{M}+\frac{p^{\prime2}}{M},\boldsymbol{q}\right)g\left(q_{0}^{\prime},\boldsymbol{q}\right)}{q_{0}^{\prime}-\frac{k^{\prime2}}{M}+\frac{p^{2}}{M}}\\
 &  & \times\langle \boldsymbol{k}|\tilde{\rho}_{o}\left(t\right)|\boldsymbol{p}^{\prime}\rangle\langle \boldsymbol{p}^{\prime}|C_{\boldsymbol{q}.\hat{\boldsymbol{s}}}|\boldsymbol{p}\rangle \langle \boldsymbol{p}|C_{\boldsymbol{q}.\hat{\boldsymbol{s}}^{\prime}}|\boldsymbol{k}^{\prime}\rangle \\
\\
 &  & -\frac{1}{2N_{c}}\underset{m}{\sum}\int_{\boldsymbol{q}\boldsymbol{p}}\text{P}\int_{q_{0}^{\prime}}\frac{g\left(q_{0}^{\prime}-\frac{p^{2}}{M}+\frac{k^{2}}{M},\boldsymbol{q}\right)g\left(q_{0}^{\prime},\boldsymbol{q}\right)}{q_{0}^{\prime}-\frac{p^{2}}{M}+E_{m}}\\
 &  & \times\langle \boldsymbol{k}|S_{\boldsymbol{q}.\hat{\boldsymbol{s}}}|m\rangle \langle m|S_{\boldsymbol{q}.\hat{\boldsymbol{s}}^{\prime}}|\boldsymbol{p}\rangle \langle \boldsymbol{p}|\tilde{\rho}_{o}\left(t\right)|\boldsymbol{k}^{\prime}\rangle \\
\\
 &  & +\frac{1}{2N_{c}}\underset{m}{\sum}\int_{\boldsymbol{q}\boldsymbol{p}}\text{P}\int_{q_{0}^{\prime}}\frac{g\left(q_{0}^{\prime}-\frac{k^{\prime2}}{M}+\frac{p^{2}}{M},\boldsymbol{q}\right)g\left(q_{0}^{\prime},\boldsymbol{q}\right)}{q_{0}^{\prime}-\frac{k^{\prime2}}{M}+E_{m}}\\
 &  & \times\langle \boldsymbol{k}|\tilde{\rho}_{o}\left(t\right)|\boldsymbol{p}\rangle \langle \boldsymbol{p}|S_{\boldsymbol{q}.\hat{\boldsymbol{s}}}|m\rangle \langle m|S_{\boldsymbol{q}.\hat{\boldsymbol{s}}^{\prime}}|\boldsymbol{k}^{\prime}\rangle  
\end{array},
\end{equation}
\begin{equation}
\begin{array}{ccl}
\langle \boldsymbol{k}|\tilde{\mathcal{L}}_{os}\left(t\right)\tilde{\rho}_{s}\left(t\right)|\boldsymbol{k}^{\prime}\rangle  & = & \frac{1}{2N_{c}}\underset{m,m^{\prime}}{\sum}\int_{\boldsymbol{q}}g\left(E_{m^{\prime}}-\frac{k^{\prime2}}{M},\boldsymbol{q}\right)g\left(E_{m}-\frac{k^{2}}{M},\boldsymbol{q}\right)\\
 &  & \times\langle \boldsymbol{k}|S_{\boldsymbol{q}.\hat{\boldsymbol{s}}^{\prime}}|m\rangle \langle m|\tilde{\rho}_{s}\left(t\right)|m^{\prime}\rangle \langle m^{\prime}|S_{\boldsymbol{q}.\hat{\boldsymbol{s}}}|\boldsymbol{k}^{\prime}\rangle 
\end{array},\label{l_os-rho_s-element}
\end{equation}
\begin{equation}
\begin{array}{ccl}
\langle \boldsymbol{k}|\tilde{\mathcal{L}}_{oo}\left(t\right)\tilde{\rho}_{o}\left(t\right)|\boldsymbol{k}^{\prime}\rangle  & = & \frac{N_{c}^{2}-4}{4N_{c}}\int_{\boldsymbol{q}\boldsymbol{p}\boldsymbol{p}^{\prime}}g\left(\frac{p^{\prime2}}{M}-\frac{k^{\prime2}}{M},\boldsymbol{q}\right)g\left(\frac{p^{2}}{M}-\frac{k^{2}}{M},\boldsymbol{q}\right)\\
 &  & \times\langle \boldsymbol{k}|S_{\boldsymbol{q}.\hat{\boldsymbol{s}}^{\prime}}|\boldsymbol{p}\rangle \langle \boldsymbol{p}|\tilde{\rho}_{o}\left(t\right)|\boldsymbol{p}^{\prime}\rangle\langle \boldsymbol{p}^{\prime}|S_{\boldsymbol{q}.\hat{\boldsymbol{s}}}|\boldsymbol{k}^{\prime}\rangle  \\
\\
 &  & +\frac{N_{c}}{4}\int_{\boldsymbol{q}\boldsymbol{p}\boldsymbol{p}^{\prime}}g\left(\frac{p^{\prime2}}{M}-\frac{k^{\prime2}}{M},\boldsymbol{q}\right)g\left(\frac{p^{2}}{M}-\frac{k^{2}}{M},\boldsymbol{q}\right)\\
 &  & \times\langle \boldsymbol{k}|C_{\boldsymbol{q}.\hat{\boldsymbol{s}}^{\prime}}|\boldsymbol{p}\rangle \langle \boldsymbol{p}|\tilde{\rho}_{o}\left(t\right)|\boldsymbol{p}^{\prime}\rangle\langle \boldsymbol{p}^{\prime}|C_{\boldsymbol{q}.\hat{\boldsymbol{s}}}|\boldsymbol{k}^{\prime}\rangle \\
\\
 &  &-\frac{1}{4N_{c}}\underset{m}{\sum}\int_{\boldsymbol{q}\boldsymbol{p}}g\left(\frac{p^{2}}{M}-E_{m},\boldsymbol{q}\right)g\left(\frac{k^{2}}{M}-E_{m},\boldsymbol{q}\right)\\
 &  & \times\langle \boldsymbol{k}|S_{\boldsymbol{q}.\hat{\boldsymbol{s}}}|m\rangle \langle m|S_{\boldsymbol{q}.\hat{\boldsymbol{s}}^{\prime}}|\boldsymbol{p}\rangle \langle \boldsymbol{p}|\tilde{\rho}_{o}\left(t\right)|\boldsymbol{k}^{\prime}\rangle \\
\\
 &  & -\frac{1}{4N_{c}}\underset{m}{\sum}\int_{\boldsymbol{q}\boldsymbol{p}}g\left(\frac{p^{\prime2}}{M}-E_{m},\boldsymbol{q}\right)g\left(\frac{k^{2}}{M}-E_{m},\boldsymbol{q}\right)\\
 &  & \times\langle \boldsymbol{k}|\tilde{\rho}_{o}\left(t\right)|\boldsymbol{p}\rangle \langle \boldsymbol{p}|S_{\boldsymbol{q}.\hat{\boldsymbol{s}}}|m\rangle \langle m|S_{\boldsymbol{q}.\hat{\boldsymbol{s}}^{\prime}}|\boldsymbol{k}^{\prime}\rangle \\
 \\
&& -\frac{N^2_{c}-4}{8N_c}\int_{\boldsymbol{q}\boldsymbol{p}\boldsymbol{p}^\prime}g\left(\frac{p^{2}}{M}-\frac{p^{\prime 2}}{M},\boldsymbol{q}\right)g\left(\frac{k^{2}}{M}-\frac{p^{\prime 2}}{M},\boldsymbol{q}\right)\\
 &  & \times\langle \boldsymbol{k}|S_{\boldsymbol{q}.\hat{\boldsymbol{s}}}|m\rangle \langle m|S_{\boldsymbol{q}.\hat{\boldsymbol{s}}^{\prime}}|\boldsymbol{p}\rangle \langle \boldsymbol{p}|\tilde{\rho}_{o}\left(t\right)|\boldsymbol{k}^{\prime}\rangle \\
\\
 &  & -\frac{N^2_{c}-4}{8N_c}\int_{\boldsymbol{q}\boldsymbol{p}\boldsymbol{p}^\prime}g\left(\frac{k^{\prime2}}{M}-\frac{p^{\prime 2}}{M},\boldsymbol{q}\right)g\left(\frac{p^{2}}{M}-\frac{p^{\prime 2}}{M},\boldsymbol{q}\right)\\
 &  & \times\langle \boldsymbol{k}|\tilde{\rho}_{o}\left(t\right)|\boldsymbol{p}\rangle \langle \boldsymbol{p}|S_{\boldsymbol{q}.\hat{\boldsymbol{s}}}|m\rangle \langle m|S_{\boldsymbol{q}.\hat{\boldsymbol{s}}^{\prime}}|\boldsymbol{k}^{\prime}\rangle \\
 \\
 &&-\frac{N_{c}}{8}\int_{\boldsymbol{q}\boldsymbol{p}\boldsymbol{p}^\prime}g\left(\frac{p^{2}}{M}-\frac{p^{\prime 2}}{M},\boldsymbol{q}\right)g\left(\frac{k^{2}}{M}-\frac{p^{\prime 2}}{M},\boldsymbol{q}\right)\\
 &  & \times\langle \boldsymbol{k}|C_{\boldsymbol{q}.\hat{\boldsymbol{s}}}|m\rangle \langle m|C_{\boldsymbol{q}.\hat{\boldsymbol{s}}^{\prime}}|\boldsymbol{p}\rangle \langle \boldsymbol{p}|\tilde{\rho}_{o}\left(t\right)|\boldsymbol{k}^{\prime}\rangle \\
\\
 &  & -\frac{N_{c}}{8}\int_{\boldsymbol{q}\boldsymbol{p}\boldsymbol{p}^\prime}g\left(\frac{k^{\prime2}}{M}-\frac{p^{\prime 2}}{M},\boldsymbol{q}\right)g\left(\frac{p^{2}}{M}-\frac{p^{\prime 2}}{M},\boldsymbol{q}\right)\\
 &  & \times\langle \boldsymbol{k}|\tilde{\rho}_{o}\left(t\right)|\boldsymbol{p}\rangle \langle \boldsymbol{p}|C_{\boldsymbol{q}.\hat{\boldsymbol{s}}}|m\rangle \langle m|C_{\boldsymbol{q}.\hat{\boldsymbol{s}}^{\prime}}|\boldsymbol{k}^{\prime}\rangle 
\\
\end{array}
\end{equation}
\section{Computation details of the semiclassical limit }\label{Appendix-computation-details-semiclassical-limit}
In this appendix, we provide some of the computation details underlying the derivation of the Boltzmann equations starting from the ULE. As an example, we show the key steps involved in implementing the semiclassical approximation for the octet-to-singlet transition element
\begin{equation}
\begin{array}{ccl}
\langle n,\boldsymbol{P}_1|{\mathcal{L}}_{so}\left(t\right){\rho_{o}}\left(t\right)|n^{\prime},\boldsymbol{P}_2\rangle  & = & C_{F}\int_{\boldsymbol{q}\boldsymbol{p}\boldsymbol{p}^{\prime}} g\left(\frac{p^{2}}{M}-E_{n^{\prime}},\boldsymbol{q}\right)g\left(\frac{p^{\prime2}}{M}-E_{n},\boldsymbol{q}\right)\\
 &  & \times\langle n|S_{\boldsymbol{q}.\hat{\boldsymbol{s}}^{\prime}}|\boldsymbol{p}^{\prime}\rangle \langle \boldsymbol{p}^{\prime},\boldsymbol{P}_1|{\rho}_{o}\left(t\right)|\boldsymbol{p},\boldsymbol{P}_2\rangle \langle \boldsymbol{p}|S_{\boldsymbol{q}.\hat{\boldsymbol{s}}}|n^{\prime}\rangle\label{194-D}
\end{array},
\end{equation}
and similar steps can be followed for the other terms.

First, as in the main text, we restrict ourselves to the diagonal elements of the density matrix and set $n^\prime=n$. Similarly, we assume a small coherence length, $\boldsymbol{y}=\boldsymbol{s}^\prime-\boldsymbol{s}\simeq0$, and set $\boldsymbol{s}^\prime=\boldsymbol{s}$. Subsequently, using Eqs.~(\ref{wigner-singlet})--(\ref{wigner-octet}), we perform the Wigner transform of both sides of Eq.~(\ref{194-D}). This yields
\begin{equation}
\begin{array}{ccl}
\frac{\partial f_n(t,\boldsymbol{R},\boldsymbol{P})}{\partial t}|_{so}  & = & C_{F}\int_{\boldsymbol{q}\boldsymbol{p}\boldsymbol{p}^{\prime}\Delta \boldsymbol{P}\boldsymbol{R}^\prime\boldsymbol{r}}e^{i\Delta\boldsymbol{P}.\boldsymbol{R}}e^{-i\Delta\boldsymbol{P}.\boldsymbol{R}^\prime}e^{-i\Delta\boldsymbol{p}.\boldsymbol{r}} g\left(\frac{p^{2}}{M}-E_{n},\boldsymbol{q}\right)g\left(\frac{p^{\prime2}}{M}-E_{n},\boldsymbol{q}\right)\\
 &  & \times\langle n|S_{\boldsymbol{q}.\hat{\boldsymbol{s}}}|\boldsymbol{p}^{\prime}\rangle\langle \boldsymbol{p}|S_{\boldsymbol{q}.\hat{\boldsymbol{s}}}|n\rangle f_o\left(t, \boldsymbol{R}^\prime,\boldsymbol{P},\boldsymbol{r},\tilde{\boldsymbol{p}}\right)\\
 \\
&&=C_{F}\int_{\boldsymbol{q}\boldsymbol{p}\boldsymbol{p}^{\prime}\boldsymbol{r}}e^{-i\Delta\boldsymbol{p}.\boldsymbol{r}} g\left(\frac{p^{2}}{M}-E_{n},\boldsymbol{q}\right)g\left(\frac{p^{\prime2}}{M}-E_{n},\boldsymbol{q}\right)\\
 &  & \times\langle n|S_{\boldsymbol{q}.\hat{\boldsymbol{s}}}|\boldsymbol{p}^{\prime}\rangle\langle \boldsymbol{p}|S_{\boldsymbol{q}.\hat{\boldsymbol{s}}}|n\rangle f_o\left(t, \boldsymbol{R},\boldsymbol{P},\boldsymbol{r},\tilde{\boldsymbol{p}}\right)
\end{array},
\end{equation}
where $(\boldsymbol{R},\boldsymbol{P})$ and $(\boldsymbol{r},\tilde{\boldsymbol{p}})$ are the center-of-mass and relative coordinates, respectively, with  $\boldsymbol{P}=\frac{\boldsymbol{P}_1+\boldsymbol{P}_2}{2}$, $\tilde{\boldsymbol{p}}=\frac{\boldsymbol{p}+\boldsymbol{p}^\prime}{2}$, $\Delta\boldsymbol{P}=\boldsymbol{P}_2-\boldsymbol{P}_1$, and $\Delta\boldsymbol{p}=\boldsymbol{p}^\prime-\boldsymbol{p}$. 
We can now consider the semiclassical, or gradient, expansion of the octet Wigner function \cite{haug2008quantum, yao2021semiclassical-transport}
\begin{equation}
f_o\left(t, \boldsymbol{R},\boldsymbol{P},\boldsymbol{r},\tilde{\boldsymbol{p}}\right)=f_o\left(t, \boldsymbol{R},\boldsymbol{P},\boldsymbol{r}_0,\tilde{\boldsymbol{p}}\right)+\left(\boldsymbol{r}-\boldsymbol{r}_0\right).\nabla_{\boldsymbol{r}_0}f_o\left(t, \boldsymbol{R},\boldsymbol{P},\boldsymbol{r}_0,\tilde{\boldsymbol{p}}\right)+\cdots.  \label{gradient-expansion} 
\end{equation}
the second- and higher-order terms account for quantum corrections to the semiclassical Boltzmann equation.
At this stage, we restrict ourselves to the leading-order term and set, for simplicity, $\boldsymbol{r}_0=0$.\footnote{This is equivalent to the assumption  of an extended octet state, as implemented in the main text.} This yields,   
\begin{equation}
\begin{array}{ccl}
\frac{\partial f_n(t,\boldsymbol{R},\boldsymbol{P})}{\partial t}|_{so}  & = & C_{F}\int_{\boldsymbol{q}\boldsymbol{p}\boldsymbol{p}^{\prime}} \delta(\Delta \boldsymbol{p}) g\left(\frac{p^{2}}{M}-E_{n},\boldsymbol{q}\right)g\left(\frac{p^{\prime2}}{M}-E_{n},\boldsymbol{q}\right)\\
 &  & \times\langle n|S_{\boldsymbol{q}.\hat{\boldsymbol{s}}}|\boldsymbol{p}^{\prime}\rangle\langle \boldsymbol{p}|S_{\boldsymbol{q}.\hat{\boldsymbol{s}}}|n\rangle f_o\left(t, \boldsymbol{R},\boldsymbol{P},\boldsymbol{0},\tilde{\boldsymbol{p}}\right)\\
 \\
 &&=C_{F}\int_{\boldsymbol{q}\boldsymbol{p}}  \left[g\left(\frac{p^{2}}{M}-E_{n},\boldsymbol{q}\right)\right]^2
|\langle n|S_{\boldsymbol{q}.\hat{\boldsymbol{s}}}|\boldsymbol{p}\rangle|^2 f_o\left(t, \boldsymbol{R},\boldsymbol{P},\boldsymbol{0},\boldsymbol{p}\right)
 \end{array}
 \end{equation}
which is identical to the expression (\ref{5.11}) given in the main text. The leading quantum correction is derived in Sec.~\ref{section-quantum-corrections}.

\bibliographystyle{unsrt}

\bibliography{biblio_cited_only}
\end{document}